\documentclass[11pt,twoside]{article}
\usepackage{tikz}
\usetikzlibrary{shapes}
\usetikzlibrary{intersections}   
\usetikzlibrary{arrows,decorations.markings}  
\usepackage[normalem]{ulem}
\usetikzlibrary{positioning}
\usetikzlibrary{arrows.meta}
\usepackage{pgfplots}
\pgfplotsset{compat=1.18}
\usepackage{amsmath,amsthm}
\usepackage{graphicx}
\usepackage{geometry}
\usepackage{multirow}
\usepackage{float}
\usepackage{sectsty}
\usepackage{titlesec}
\usepackage{amssymb}

\usepackage[utf8]{inputenc}
\usepackage[T1]{fontenc}
\usepackage{lmodern}
\usepackage[english]{babel}
\usepackage{braket}
\usepackage{mathrsfs}
\usepackage{titleps}
\usepackage{tabularx}
\usepackage{enumitem}
\usepackage{hyperref}
\usepackage{wrapfig}
\usepackage{color}
\usepackage[perpage]{footmisc}
\usepackage[babel]{csquotes}
\usepackage{chngcntr}
\usepackage[version=4]{mhchem}

\newcommand{\field}[1]{\mathbb{#1}}
\newcommand{\N}{\field{N}}
\newcommand{\R}{\field{R}}
\newcommand{\C}{\field{C}}
\newcommand{\Z}{\field{Z}}

\newcommand{\Ea}{E_{\textup{at}}}
\newcommand{\Ha}{H_{\textup{at}}}

\newtheorem{theorem}{Theorem}[section]
\newtheorem{lemma}[theorem]{Lemma}
\newtheorem{remark}[theorem]{Remark}
\newtheorem{proposition}[theorem]{Proposition}
\newtheorem{defi}[theorem]{Definition}
\newtheorem{kor}[theorem]{Corollary}

\newcolumntype{L}[1]{>{\raggedright\arraybackslash}p{#1}}
\newcolumntype{C}[1]{>{\centering\arraybackslash}p{#1}}
\newcolumntype{R}[1]{>{\raggedleft\arraybackslash}p{#1}}

\titleformat{\paragraph}
{\normalfont\normalsize\bfseries}{\theparagraph}{1em}{}
\titlespacing*{\paragraph}
{0pt}{3.25ex plus 1ex minus .2ex}{1.5ex plus .2ex}



\addto{\captionsenglish}{
    
}

\sectionfont{\Large}
\subsectionfont{\Large}
\date{ }

\sectionfont{\Large}
\subsectionfont{\large}

\counterwithout{figure}{figurecount}
\begin{document}

\title{Energy and Radius Estimates for Molecules}

\author{Ioannis Anapolitanos\footnote{
Institute for Analysis, Karlsruhe Institute of Technology (KIT), D-76128 Karlsruhe, Germany} \qquad \qquad Michael Hofacker\footnotemark[1]}

\maketitle

\setcounter{section}{0}
\begin{abstract}
 The question of whether ionizing a system of well-separated atoms requires energy has remained open for decades, even though the answer is, from a physical point of view, obviously positive. The analog question for a system of molecules is even more challenging, partly because the molecular geometry is not known a priori and may change when electrons are added or removed. In this work, we investigate and analytically solve the first problem in the special case of a mixture of hydrogen and helium atoms. As a consequence, the van der Waals law for a system of hydrogen and helium atoms, which was proven in \cite{AnapolitanosSigal2013} assuming our result, now follows as a corollary without assumptions.  Moreover, we prove a new lower bound for the `size' of a stable molecule with several nuclei, which improves upon the lower bounds from \cite{MoleculesBenguria} with even less assumptions. Using among others this lower bound,  we prove that it costs energy to ionize a system consisting of an $H_2$ molecule and a perfectly conducting half-infinite plate, provided that they are well-separated. The long-range asymptotics of their interaction energy was analyzed in \cite{ABH} assuming our result. 
\end{abstract}

\section{Main results and notation}
We consider a molecule composed of $N \in \N$ electrons and $K \in \N$ nuclei with charges $\underline{Z}=(Z_1,\ldots,Z_K)$ in the Born-Oppenheimer approximation.
In appropriate units, the associated Hamiltonian is given by
\begin{align}
H(N,\underline{Z}, \underline{R}):=\;\sum_{i=1}^{N}\left( -\Delta_{x_i} - \sum_{j=1}^K \frac{Z_j}{\left|x_i-R_j\right|}\right) +  \sum_{1 \leq i< j \leq N} \frac{1}{\left|x_i-x_j\right|}
+\sum_{1 \leq i< j \leq K} \frac{Z_i Z_j}{\left|R_i-R_j\right|} \,, \label{DefHamiltonian}
\end{align}
where $\underline{R}=(R_1,\ldots,R_K)$ collects the positions of the nuclei that are considered as fixed. 
In the atomic case $(K=1)$, the Hamiltonian from Eq.  \eqref{DefHamiltonian} can be reduced to $\Ha(N,Z, 1)$, where
\begin{align}
\Ha(N,Z, \eta):=\;\sum_{i=1}^{N}\left( -\Delta_{x_i} - \frac{Z}{\left|x_i\right|}\right) +  \sum_{1 \leq i< j \leq N} \frac{\eta}{\left|x_i-x_j\right|}\,. \label{DefHat}
\end{align}
For later mathematical purposes, we have weighted the electron-electron repulsion in Eq. \eqref{DefHat} with a factor $\eta \geq 0$.  
In nature, as a consequence of the Pauli exclusion principle and the electron spin, both $H(N,\underline{Z}, \underline{R})$ 
and $\Ha(N,Z, \eta)$ act in the antisymmetric Hilbert space
\begin{align*}
\mathcal{H} = \bigwedge_{i=1}^{N} (L^2(\R^3) \otimes \C^2)\,.
\end{align*}
It is well-known that $H(N,\underline{Z}, \underline{R})$ and $\Ha(N,Z, \eta)$ define self-adjoint and semibounded operators on the domain $D(H(N,\underline{Z}, \underline{R}))=D(\Ha(N,Z, \eta))=\mathcal{H}^2$, where 
\begin{align*}
\mathcal{H}^2:= \left\{ \psi \in \mathcal{H} \,: \psi(\, \cdot \,, s_1,\ldots, \, \cdot \,, s_N) \in H^2(\R^{3N}) \, \quad  \forall s_1,\ldots,s_N \in \left\{1,2 \right\} \right\} 
\end{align*}
denotes the subspace of $\mathcal{H}$ that contains all wave functions that lie in the usual Sobolev space $H^2(\R^{3N})$ if the spin variables are fixed. In particular, the associated ground-state energies 
\begin{align}
E(N,\underline{Z},\underline{R})&:= \inf \sigma(H(N,\underline{Z}, \underline{R})) \,,\label{DefE} \\
\Ea(N,Z,\eta)&:= \inf \sigma(\Ha(N,Z, \eta)) \label{DefEat}
\end{align}
are finite. We note that the atomic ground-state energy $\Ea(N,Z,\eta)$ is always negative, whereas this is not the case for the molecular ground-state energy $E(N,\underline{Z},\underline{R})$ with $K \geq 2$ if at least two of the nuclei are close to each other. In the physically relevant case $\eta=1$ we  set for brevity
\begin{align}
\Ha(N,Z):=\Ha(N,Z, 1) \qquad \text{and} \qquad  \Ea(N,Z):=\Ea(N,Z,1)\,. \label{Eateta=1}
\end{align} 
The molecular ground-state energy $E(N,\underline{Z},\underline{R})$ depends heavily on the positions $\underline{R}$ of the nuclei. 
However, instead of considering the total value of the ground-state energy, it is often more convenient to consider the binding energy, sometimes also called interaction energy, 
\begin{align}
W(\underline{R}):= E(N,\underline{Z}, \underline{R}) - E(N,\underline{Z},\infty)\,, \label{DefW}
\end{align}
where $E(N,\underline{Z}, \underline{R})$ denotes the molecular ground-state energy defined by Eq. \eqref{DefE} and 
\begin{align}
E(N,\underline{Z},\infty):= \lim_{\lambda \to \infty} \inf\left\{E(N,\underline{Z},\underline{R}) \,:\, \min_{i \neq j} |R_i-R_j| > \lambda \right\} \label{DefEinfinity}    
\end{align}
denotes the ground-state energy of the system when the nuclei are infinitely far from each other. The inequality $W(\underline{R})<0$ roughly means that it costs energy to separate the system completely, which is often interpreted as a sign of attraction.
In \cite{MorganSimon} it has been shown that 
\begin{align}
E(N,\underline{Z},\infty)= \min_{(N_1,\ldots,N_K)} \sum_{i=1}^{K} \Ea(N_i,Z_i)\,,\label{EqEinfinity}
\end{align}
where the minimum on the right-hand side is taken over all possible distributions $(N_1,\ldots,N_K)$ of the $N$ electrons to the $K$ nuclei (that is $N_i \in \N_0$, $i=1,\ldots,K$, and $\sum_{i=1}^K N_i=N$). 
If the system is neutral in total, i.e. $N= \sum_{i=1}^{K} Z_i$, then one expects that the minimum in Eq. \eqref{EqEinfinity} is achieved if the system is decomposed into neutral atoms, i.e. $N_i=Z_i$ for $i=1,\ldots,K$. More precisely, if $n_i \in \Z$ and $N_i+n_i \geq 0$ for $i=1,...,K$ with $\sum\limits_{i=1}^K n_i=0$ and $\sum\limits_{i=1}^K |n_i| \neq 0$, then one expects that
\begin{align*}
	\sum_{i=1}^K \Ea(N_i,N_i) < \sum_{i=1}^K \Ea(N_i+n_i,N_i). \qquad \qquad (\text{E}\,)
\end{align*}
The latter inequality is referred to as Condition (E) in the following. Its interpretation is that is costs energy to ionize a system of well-separated atoms.
In fact, Condition (E) is a necessary condition for the van der Waals law described below \cite{AnapolitanosSigal2013}. 
 In the diatomic case ($K=2$), the binding energy between two  atoms is often described by the Lennard-Jones potential $W_{\textup{LJ}}(R)$ (see Figure \ref{Fig:LennardJones} below).
Due to the repulsion between the nuclei, 
$W_{\textup{LJ}}(R)$ is strongly repulsive for small distances $R$ between the nuclei (a common choice is $W_{\textup{LJ}}(R) \sim 1/R^{12}$ for small $R$), whereas for large distances $W_{\textup{LJ}}(R)$ is attractive with $W_{\textup{LJ}}(R) \sim -1/R^{6}$ due to the van der Waals force between the two atoms. The minimum of the ground-state energy is attained at some
intermediate distance $R_0$, which can be interpreted as the binding length of the molecule.
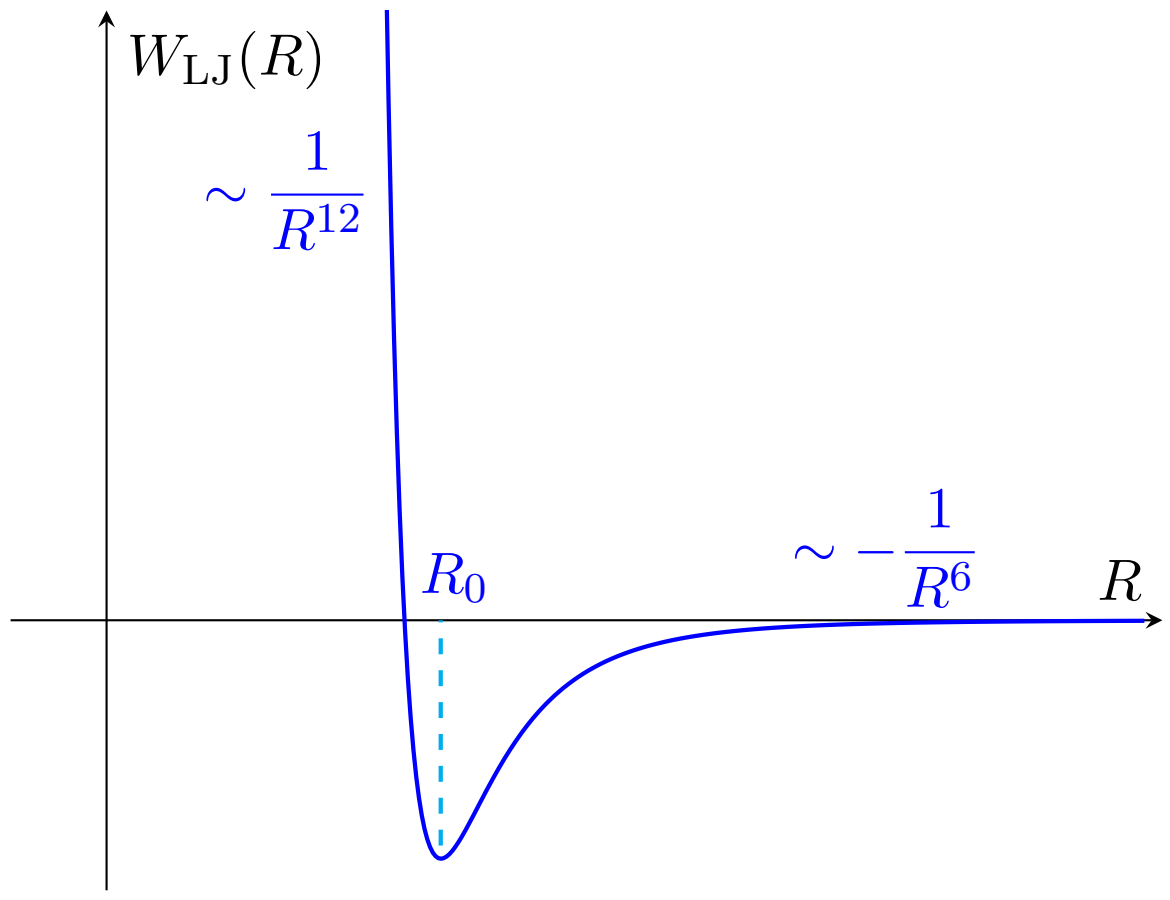
\begin{figure}[H] 
\begin{center}
\begin{tikzpicture}[scale=1]
\begin{axis}[width=9.2cm, height=7.4cm,
    trig format plots=rad,
    samples=300,
    %axis equal image,
   xlabel=$R$,ylabel=$W_{\textup{LJ}}(R)$,
    axis lines=middle, enlargelimits=true, color=black,
    xmin=0, xmax=1.6, ymin=-0.55, ymax=1.5, xtick=\empty,
 ytick= \empty
   ]
\node[color=blue, font=\normalsize] at (axis cs: 0.3,1.2) {$\sim \dfrac{1}{R^{12}}$};
\node[color=blue, font=\normalsize] at (axis cs: 1.3,0.2) {$\sim -\dfrac{1}{R^{6}}$};
\node[color=blue, font=\normalsize] at (axis cs: 0.58,0.12) {$R_0$};

\addplot [
    no markers,
    blue,
    thick,
    domain=0.35:1.73
] {0.0006*x^(-12)-0.04*x^(-6)};

 \addplot [mark=none, dashed, thick, color=cyan] coordinates {(0.557, -0.63) (0.557, 0)};
\end{axis}
\end{tikzpicture}
\caption{\label{Fig:LennardJones} Lennard-Jones potential.}
\end{center}
\end{figure}
For systems with more than two nuclei, similar considerations can be made. In previous works like \cite{AnapolitanosSigal2013}
 the binding energy $W(\underline{R})$ for large distances 
\begin{align*}
R:=\min_{i=1,\ldots,K} |R_i-R_j|
\end{align*}
between the nuclei was investigated. More precisely, assuming among others Condition (E), the van der Waals law 
\begin{align}
W(\underline{R})= \sum_{1 \leq i < j \leq K} -\dfrac{\sigma_{ij}}{|R_i-R_j|^6} + o\left( \dfrac{1}{R^6} \right) \qquad \qquad (R \to \infty) \label{VanderWaalslaw}
\end{align}
was established for several atoms.
 Here, $\sigma_{ij}$, $1\leq i<j \leq K$, are positive coefficients  depending on  $\underline{Z}$, but not on $\underline{R}$. The remainder $o\left(\ldots\right)$ might also depend on $N$ and $\underline{Z}$. 
 Eq. \eqref{VanderWaalslaw} implies that 
 $W(\underline{R})$ is negative for large $R$, which 
 in view of Eq. \eqref{DefW} means that it costs energy to separate the system. Prior to \cite{AnapolitanosSigal2013}, an upper bound for the interaction energy for several molecules was proven in \cite{LiebThirringvdW}, confirming the existence of the van der Waals attraction.  After \cite{AnapolitanosSigal2013}, several extensions  establishing the van der Waals law appeared in the literature. In \cite{AnapLew}, an analog of Eq. \eqref{VanderWaalslaw} was proven for a system of several molecules, and in the case of two molecules it was applied to show boundedness of reaction paths for a special type of isomerizations. 
 In \cite{BaHaHuVu}, the expansion of the interaction energy also included higher 
order terms than \eqref{VanderWaalslaw}, with remainder
$O\left(1/R^{10}\right)$,
and electrons with semi-relativistic kinetic energy were considered as well. Using among others methods of  \cite{AnapLew} and \cite{BaHaHuVu}, the results of  \cite{AnapLew} were extended in \cite{AnOlZa} to systems with electrons having semi-relativistic kinetic energy.   Moreover, the works \cite{CaCoSc, CaSc} approximated numerically the coefficient $\sigma_{12}$ in \eqref{VanderWaalslaw} as well as some higher order coefficients in the special case of two hydrogen atoms. 
It should be noted that Eq. \eqref{VanderWaalslaw} does not suffice to establish attraction between atoms of a real gas. An important reason for this is that the remainder in Eq. \eqref{VanderWaalslaw}, or in analogs thereof, depends on the number of atoms involved, and it is not clear how. To prove attraction for real gases, one has to study the interaction energy in the joint limit $R \rightarrow \infty, K \rightarrow \infty$, where we recall that $K$ is the number of nuclei. To the best of our knowledge, the only work that considers this joint limit is \cite{Anap16}. However, the remainder estimates obtained therein are far from establishing attraction for real gases. We note that Condition (E) played an important role in most of the works mentioned in this paragraph.
 
 The first goal of this work is to prove Condition (E) for  a mixed system of hydrogen and helium atoms. In particular, the van der Waals law \eqref{VanderWaalslaw} for the case of a system of helium atoms follows as a corollary with no further assumptions from our Theorem \ref{theohydrogenhelium} below and from \cite{AnapolitanosSigal2013}.
 
In the second part of this work, we provide a lower bound on the binding length $R_0$ of a diatomic molecule and, more generally, on the 'size' of a polyatomic molecule. 
 More specifically, we derive lower bounds on the 'size' of all local mimima of the molecular ground-state energy in terms of the corresponding atomic and ionic ground-state energies. The main result of this part is Theorem \ref{theoRbbound2} below. In particular, we show that our result provides stronger estimates than \cite[Theorems 1,2]{MoleculesBenguria} under weaker assumptions.
Our motivation for these lower bounds is twofold. On the one hand, the van der Waals law provides information about the long-range behavior of the interaction energy, but it does not provide any information about stable molecular configurations. On the other hand, information about $R_0$, or more generally about the 'size' of a polyatomic molecule, could be a useful tool towards proving the analog of Condition (E) for systems of molecules. This is demonstrated in the third part of our work, where in Theorem \ref{Thm:EnergyCondition} we use the lower bound on $R_0$ for the hydrogen molecular ion $\ce{H2+}$ to prove the analog of Condition (E) for a hydrogen molecule interacting with a perfectly conducting half-infinite plate.
As a corollary, the long-range asymptotics of their interaction energy, which was established in \cite{ABH} assuming our result, follows without additional assumptions.

We start by considering the problem of ionizing a system of well-separated atoms. To this end, we consider a system composed of $N$ electrons and $K \geq 2$ nuclei with charges $\underline{Z}=(Z_1,\ldots,Z_K) \in \N^K$ that is neutral in total i.e. $N= \sum_{i=1}^{K} Z_i$. 
In \cite[Theorem 1.2]{AnapolitanosSigal2013} the analog of the van der Waals law \eqref{VanderWaalslaw} has been established for fixed symmetry types under Condition (E) and another Condition called (D). Condition (D), which, roughly speaking, can be viewed as a generalized non-degeneracy condition for the ground-state subspace in the given symmetry type, is true for a system of hydrogen and helium atoms, see the discussion after Remark 1.2 below. For the precise definition of Condition (D), we refer to  \cite{AnapolitanosSigal2013}. 
We now focus on Condition (E). Although for our purposes Condition (E) will always be sufficient, we also discuss the following stronger Condition (E') that implies Condition (E): 
For all $i \neq j$, $m,n \in \N_0$ with 
$n+1 \leq N_j$, 
\begin{align*}
\Ea(N_i+m,N_i)+\Ea(N_j-n,N_j) < \Ea(N_i+m+1,N_i) + \Ea(N_j-n-1,N_j). \qquad \qquad (\text{E'}\,)
\end{align*}
Condition (E') roughly says that it costs energy to transfer an electron from a nonnegative ion to a nonpositive ion. 
We emphasize that in the case of two identical atoms Condition (E') is closely related to a famous open problem in quantum chemistry: the monotonicity of the ionization energies and electron affinities in the sense that for all $n \leq N-1$,
\begin{align*}
\Ea(N-n,N) - \Ea(N-n+1,N) \leq \Ea(N-n-1,N) - \Ea(N-n,N)
\end{align*}
with equality only if the right-hand side is zero.
Here, for $n \in \N$, the difference $$\Ea(N-n,N) - \Ea(N-n+1,N)$$  corresponds to the $n$-th ionization energy of the atom with nuclear charge number $N$, and for  $n \leq 0$ it corresponds to the $(|n|+1)$-th electron affinity, namely the energy that one gains by adding an electron to an $|n|$-times negatively charged ion. Therefore, Condition (E') for two identical atoms is equivalent to the statement that ionization energies are always larger than electron affinities. 
Although one expects that monotonicity of ionization energies and electron affinities (or, equivalently, the convexity of the ground-state energy with respect to the number of electrons) holds for atoms, the situation might be different for molecules, at least under certain conditions. In fact, Di Marino, Lewin and Nenna  \cite{LewinConvexityN} have recently given a counterexample of a molecule composed of six nuclei with small fractional charges for which the ground-state energy is not convex in the number of electrons. 
Condition (E) for several hydrogen atoms was already known and is merely an exercise, in fact it follows from Lemma \ref{LemmaEN} below.  Therefore, the van der Waals law for a gas of hydrogen atoms follows as a corollary from \cite[Theorem 1.2]{AnapolitanosSigal2013}, as noted there. Our first result shows that  Condition (E) also holds for any mixture of hydrogen and helium atoms:

\begin{theorem} \label{theohydrogenhelium}
If $N_i \in \left\{1,2\right\}$ and $n_i \in \Z$ with $N_i+n_i \geq 0$ for $i=1,...,K$, $\sum\limits_{i=1}^K n_i=0$ and $\sum\limits_{i=1}^K |n_i| \neq 0$, then
\begin{align*}
\sum_{i=1}^K \Ea(N_i,N_i) < \sum_{i=1}^K \Ea(N_i+n_i,N_i)\,.
\end{align*}
\end{theorem}

\begin{remark} In the special case of two helium atoms ($K=2$ and $N_1=N_2=2$) Condition (E) follows from the main theorem in \cite{Briet} and from the known Lemma \ref{LemmaEN} below. However, to the best of our knowledge, this was not the case for a system of several helium atoms, namely for $K>2$ this was unknown even if $N_i=2$ for all $i=1,\dots, K$.  We also note that the gap that is provided by our method in the case $K=2$  is better than the one in \cite{Briet}, see Remark \ref{BrietImprovement} below for more details.
\end{remark}
As mentioned above, in \cite{AnapolitanosSigal2013} an additional Condition (D) was assumed for the derivation of the van der Waals law. In case of a mixture of hydrogen and helium atoms, Condition (D) follows from Perron-Frobenius theory, as already noted in \cite{AnapolitanosSigal2013}. Therefore, Theorem \ref{theohydrogenhelium} allows us to remove all assumptions in \cite[Theorem 1.2]{AnapolitanosSigal2013} in the special case of a mixture of hydrogen and helium atoms:

\begin{kor}\label{korvdW}
Let $K \geq 2$, $\underline{Z}=(Z_1,\ldots,Z_K) \in \left\{1,2\right\}^K$ and  $N=\sum_{i=1}^{K} Z_i$. Then the  van der Waals law \eqref{VanderWaalslaw} holds for $W(\underline{R})$ defined by Eq. \eqref{DefW}. 
\end{kor}

The van der Waals law \eqref{VanderWaalslaw} establishes the long-range behavior of the binding energy $W(\underline{R})$ of a neutral system.  However, it makes no statement about the shape of minimizing configurations $\underline{R_0} \in \R^{3K}$ of the ground-state energy.
Without the neutrality assumption, even if such a configuration $\underline{R_0}$ exists, then it is still not clear whether there exists a corresponding normalized ground state. Moreover, it is not clear whether it is stable in the sense that there is no sequence of configurations leading to the same binding energy, while the size of the corresponding molecules diverges. Therefore, we recall the following definition (see e.g. \cite{MoleculesBenguria}):

\begin{defi}\label{Defstable}
Let $K \geq 2$, $N \in \N$ and $\underline{Z}=(Z_1,\ldots,Z_K) \in (0,\infty)^K$ be given. We say that $\underline{R}_0$ is a stable configuration  for the Hamiltonian $H(N,\underline{Z}, \,\cdot\,)$ from Eq. \eqref{DefHamiltonian} if the following two conditions are satisfied:
\begin{enumerate}
\item[\bf i)] 
The function $\underline{R} \mapsto E(N,\underline{Z},\underline{R})$ attains its  global minimum at $\underline{R}=\underline{R_0}$, i.e. 
$E(N,\underline{Z},\underline{R}_0)=E(N,\underline{Z})$, where
\begin{align}
E(N,\underline{Z}):=\inf_{\underline{R}} E(N,\underline{Z},\underline{R})   \label{DefEabs}
\end{align}
and, moreover, $E(N,\underline{Z},\underline{R}_0)$ is a discrete eigenvalue of $H(N,\underline{Z}, \underline{R}_0)$.
\item[\bf ii)] It costs energy to break the system into at least two parts, more precisely $$E(N,\underline{Z}) < \lim_{\lambda \to \infty} \inf\left\{E(N,\underline{Z},\underline{R}) \, | \, \max_{i \neq j} |R_i-R_j| > \lambda \right\}.$$
\end{enumerate}
If $\underline{R}_0$ is a stable configuration, then we define the `size' of the corresponding molecule as
\begin{align}
R_b:= \max_{i \neq j} | R_{0,i} - R_{0,j}|\,. \label{DefRb}
\end{align}
\end{defi} \medskip

\begin{remark} For two neutral atoms that interact with each other, Condition (E) implies the existence of a configuration that is stable in the sense of Definition \ref{Defstable}. This follows from the upper bound on the interaction energy $W(R)$ derived in \cite{LiebThirringvdW}
in combination with the asymptotics
$W(R) \to \infty$ as $R \to 0$ and $W(R) \to 0$ as $R \to \infty$ because $W(R)$ depends continuously
 on the distance $R$ between the nuclei. If Condition (E) does not hold, then stability still holds because oppositely charged ions attract each other with Coulomb attraction. We refer e.g. to the end of Section 3 in \cite{AnapolitanosSigal2013} for a proof of the Coulomb attraction.
 \end{remark}

In \cite[THEOREM 1]{MoleculesBenguria}, it has been shown that for any stable configuration $\underline{R}_0$,
\begin{align}
R_b \geq \dfrac{Z^2- \sum_{i=1}^K (Z_i)^2}{2\sigma_N\left(
Z^2- \frac{1}{K} \sum_{i=1}^K (Z_i)^2 \right)}\,, \label{BenguriaEestimate}
\end{align}
where $Z:=\sum_{i=1}^K Z_i$ and, for brevity,
\begin{align}
\sigma_N:= |\Ea(N,1,0)| \label{DefSigmaN}
\end{align}
with $\Ea(N,1,0)$ defined by Eq. \eqref{DefEat}. Note that 
$\sigma_N$  is just the sum of the absolute values of the lowest $N$ eigenvalues of the hydrogen atom counted with multiplicity. 

Our next result gives a lower bound for the `size' of  local minima $\underline{R}_0$ of the ground-state energy.
The proof is based on a lemma of Thirring and the united atom bound, which is also used in \cite{MoleculesBenguria} to establish Estimate \eqref{BenguriaEestimate}. 
As a corollary, we obtain the lower bounds from Estimates \eqref{Rbnewestimate} and \eqref{Rbestimate} below. If $N \geq 2$, then all these estimates 
improve upon Estimate \eqref{BenguriaEestimate}, see Remark \ref{improvement} and, in particular, Estimate \eqref{Rbestimateexplicit} below. Furthermore, our proof works for all local minima of the ground-state energy, which is a weaker assumption than being a stable configuration in the sense of Definition \ref{Defstable}.

\begin{theorem} \label{theoRbbound2}
 Let $K \geq 2$, $N \in \N$, $\underline{Z} \in (0,\infty)^K$ and let $R_b$ be defined as in Eq. \eqref{DefRb} for some configuration $\underline{R}_0$ of $H(N,\underline{Z},\underline{R})$ that is a local minimum of the ground-state energy $E(N,\underline{Z},\underline{R})$. Let also $Z:=\sum_{i=1}^K Z_i$. Then $\Ea(N,Z)<E(N,\underline{Z},\underline{R}_0)<0$ and
%Then, $E(N,\underline{Z}, \lambda\underline{R}_0) \leq 0$ for all $\lambda \geq 1/2$ \textcolor{red}{Das stimmt aber folgt nicht sofort aus der Gleichung \eqref{TangentUpper}. Verwenden wir tatsächlich auch nur für $\lambda =\frac{1}{2}$. Sollen wir es trotzdem für $\lambda \geq 1/2$ erklären?}. Moreover, with $Z:=\sum_{i=1}^K Z_i$ we have that
\begin{align}\label{Rbglobalestimate}
R_b \geq \dfrac{Z^2 - \sum_{i=1}^K (Z_i)^2}{4\sqrt{|E(N,\underline{Z},\underline{R}_0)|}\left(\sqrt{|\Ea(N,Z)|}-\sqrt{|E(N,\underline{Z},\underline{R}_0)|}\right)\,}.
\end{align}
\end{theorem}
\begin{kor}
Under the assumptions of Theorem \ref{theoRbbound2}, we have $\Ea(N,Z)<E(N,\underline{Z},\underline{R}_0)<0$ and Estimate \eqref{Rbglobalestimate} can be rewritten in the form 
\begin{align}\label{Rbnewestimate}
R_b \geq \dfrac{Z^2 - \sum_{i=1}^K (Z_i)^2}{ 2 \left( E(N,\underline{Z},\underline{R}_0)-\Ea(N,Z)- \left(\sqrt{|E(N,\underline{Z},\underline{R}_0)|}-\sqrt{|\Ea(N,Z)|}\right)^2\, \right)}\,.
\end{align}
In particular, this implies that
\begin{align}\label{Rbestimate}
R_b \geq \dfrac{Z^2 - \sum_{i=1}^K (Z_i)^2}{|\Ea(N,Z)|}\,.
\end{align}
\end{kor}
\begin{proof}
    Estimate \eqref{Rbnewestimate} follows from Estimate \eqref{Rbglobalestimate} and the equality $4b(a-b) = 2(a^2-b^2-(b-a)^2)$. Estimate \eqref{Rbestimate} follows from Estimate \eqref{Rbglobalestimate} and the inequality $4b(a-b) \leq a^2$.
\end{proof}

\begin{remark}\label{improvement}
By the monotonicity of $\Ea(N,Z,\eta)$ w.r.t. $\eta$ and by scaling, we have the rough lower bound 
\begin{align}
\Ea(N,Z)=\Ea(N,Z,1) \geq  \Ea(N,Z,0) = -\sigma_N Z^2\label{Elowertrivial}
\end{align} 
with $\sigma_N$ defined by Eq. \eqref{DefSigmaN}. Note that equality  holds in Eq. \eqref{Elowertrivial} if and only if $N=1$ because for $N \geq 2$ we even have strict monotonicity of $\Ea(N,Z,\eta)$  w.r.t. $\eta$. Hence, we obtain from Estimate \eqref{Rbestimate} and from the known inequality $\Ea(N,Z)<0$ the explicit estimate 
\begin{align}
R_b \geq \dfrac{Z^2 - \sum_{i=1}^K (Z_i)^2}{\sigma_N Z^2}\,. \label{Rbestimateexplicit}
\end{align}
Note that this estimate improves Estimate \eqref{BenguriaEestimate} since
\begin{align*}
2\left( Z^2- \frac{1}{K} \sum_{i=1}^K (Z_i)^2 \right) \geq 2Z^2- \sum_{i=1}^K (Z_i)^2 > Z^2\,.
\end{align*}
Estimate \eqref{Rbestimateexplicit} can be further improved for sufficiently large $N \in \N$ by using instead of Estimate \eqref{Elowertrivial} the lower bound from \cite[Corollary 2.3]{LowerBoundsAtoms}, which in our units is given by
\begin{align*}
\Ea(N,Z) \geq -\frac{1}{2}\max((C_1 Z -C_2N)^2, C_3 Z^2) \,N^{1/3}  
\end{align*}
with the constants
$C_1=(3/2)^{1/6} \approx 1.07$, 
$C_2= ((3/2)^{1/6} - 2^{5/6} 3^{7/6}/7)/1.1185 \approx 0.1367$
and $C_3= (2^{5/6} 3^{7/6}/7)^2\approx 0.8410$.
Using this to estimate the right-hand side of Estimate \eqref{Rbestimate}  leads to
\begin{align*}
R_b \geq \dfrac{2 \left( Z^2 - \sum_{i=1}^K (Z_i)^2 \right)}{\max((C_1 Z-C_2N)^2, C_3 Z^2) N^{1/3}}\,. 
\end{align*}
In particular, we obtain in the neutral case $N=Z$ that
\begin{align*}
R_b \geq \dfrac{2}{(C_1-C_2)^2} \frac{ Z^2 - \sum_{i=1}^K (Z_i)^2 }{Z^{7/3}} \approx 2.29627 \mspace{2mu} \frac{ Z^2 - \sum_{i=1}^K (Z_i)^2 }{Z^{7/3}} \qquad  \qquad (N=Z \in \N)\,. 
\end{align*}
Note that this estimate improves Estimate \eqref{Rbestimateexplicit} in the neutral case $N=Z \in \N$ if and only if $\sigma_N > (C_1-C_2)^2 N^{1/3}/2$, which can be shown to be equivalent to $N \geq 8$.
\end{remark}
\begin{remark}
 In the diatomic case $K=2$, another lower bound for $R_b$ was proven in  \cite[THEOREM 2]{MoleculesBenguria}, namely
\begin{align}
R_b \geq \dfrac{Z_1 Z_2}{\sigma_N(Z_1)^2 + \sigma_N(Z_2)^2}\,. \label{BenguriaEestimateDiatomic}
\end{align}
However, note that the diatomic version of Estimate \eqref{Rbestimateexplicit}, i.e.
\begin{align}
R_b \geq \dfrac{2 Z_1 Z_2}{\sigma_N (Z_1+Z_2)^2}\,,\label{Rbestimatediatomicexplicit}
\end{align}
also improves Estimate \eqref{BenguriaEestimateDiatomic} if $Z_1 \neq Z_2$. 
Furthermore, if one has a better lower bound for $\Ea(N,Z_1+Z_2)$ than Estimate \eqref{Elowertrivial}, then an even better lower bound for $R_b$ follows from the diatomic version of Estimate \eqref{Rbestimate}, i.e.
\begin{align}
R_b \geq \dfrac{2 Z_1 Z_2}{|\Ea(N,Z_1+Z_2)|}\,. \label{Rbestimatediatomic}
\end{align}
However, if the nuclear charges $Z_1, Z_2 >0$ are very asymmetrically distributed, then Estimates \eqref{Rbestimatediatomicexplicit} and \eqref{Rbestimatediatomic} are not the best known lower bounds for $R_b$:
Using a localization argument, Solovej \cite{SolovejMolecules} has shown that independently of the nuclear charges $Z_1, Z_2 >0$,
\begin{align}
R_b \geq \dfrac{1}{3\sigma_N}\,. \label{SolovejBound}
\end{align} 
Note that in contrast to Estimate \eqref{SolovejBound} the right-hand sides of Estimates \eqref{Rbestimatediatomicexplicit} and \eqref{Rbestimatediatomic} both vanish in the limit $Z_1/Z_2 \to 0$.
This means that Estimates \eqref{Rbestimatediatomicexplicit} and \eqref{Rbestimatediatomic}  are better than Estimate \eqref{SolovejBound} only if the nuclear charges do not differ too much. In particular, Estimate \eqref{Rbestimatediatomicexplicit} improves Estimate \eqref{SolovejBound} if and only if $2-\sqrt{3} < Z_1/Z_2 < 2+\sqrt{3}$.
\end{remark}
\vspace*{8mm}

We now consider the problem of an $\ce{H2}$ molecule that is interacting with a perfectly conducting half-infinite plate. Such a problem has occurred in the physics literature, see e.g. \cite{Cornu_2009}, and it was recently studied rigorously in  \cite{ABH}. For simplicity of this work, we are not going to recall  the general result from \cite[Theorem 2.3]{ABH}. Instead we are going to briefly discuss only the special case of the hydrogen molecule $\ce{H2}$ (see Figure \ref{Fig:PlateWithHydrogen} below), which is centered at zero. We consider a perfectly conducting plate
 and we assume that its height, width and thickness are infinite and that it is placed  vertically to the unit vector $v \in \R^3$ passing through $-rv$ for some $r>0$. Of course, we also assume that all nuclei are placed on the other side of the plate. 
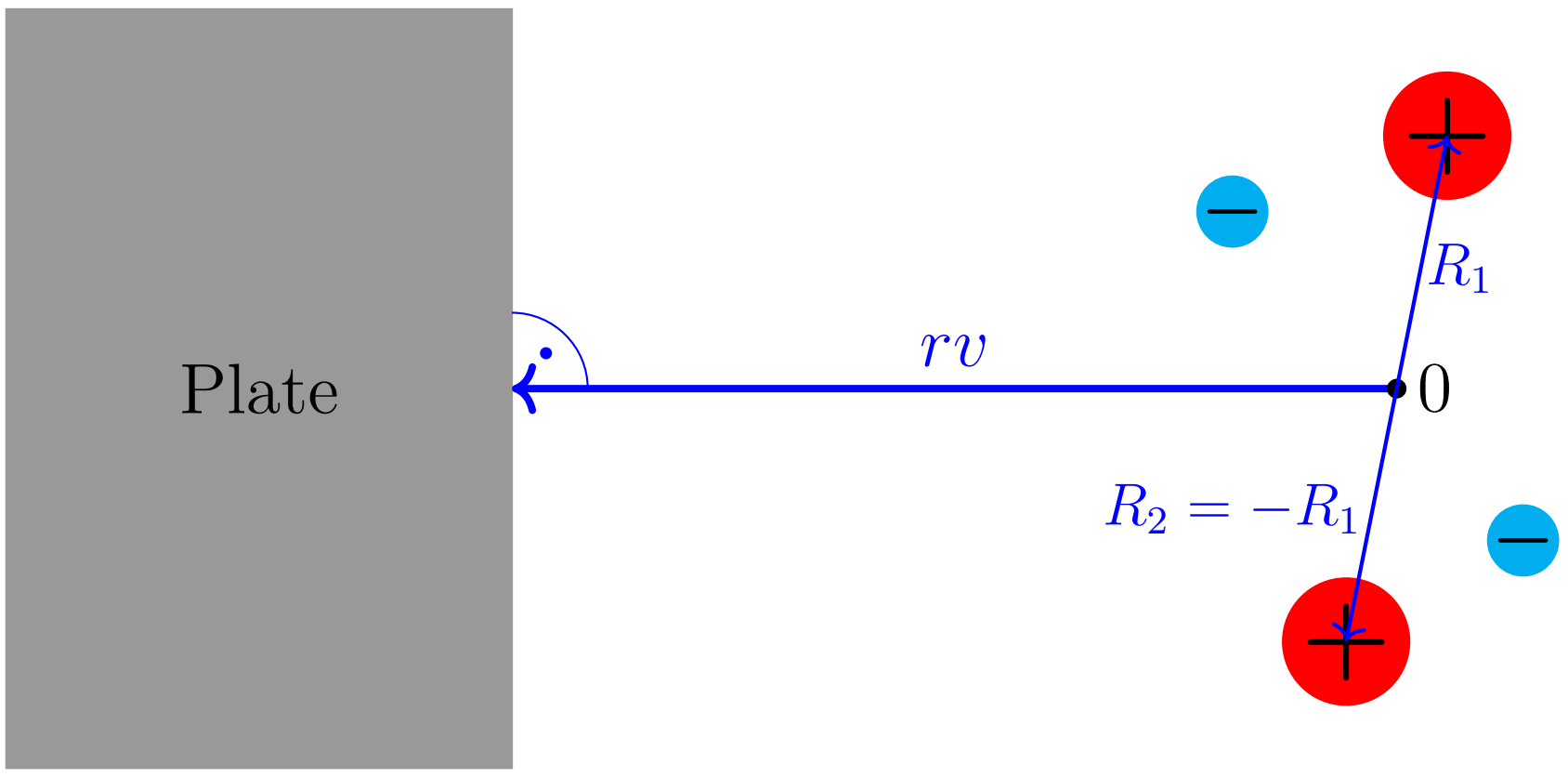
\begin{figure}[H] 
\begin{center}
\begin{tikzpicture}[scale=0.9]
\filldraw[black!40!white] (-11.0,-3.0) rectangle(-7,3.0);
\filldraw[red] (0.4,2.0) circle(0.5);
\filldraw[red] (-0.4,-2.0) circle(0.5);
\node at (0.4,2.0) {\huge{$\boldsymbol{+}$}};
\node at (-0.4,-2.0) {\huge{$\boldsymbol{+}$}};
\filldraw[cyan] (-1.3,1.4) circle(0.2778);
\filldraw[cyan] (1.0,-1.2) circle(0.2778);
\node at (-1.3,1.4) {\Large{$\boldsymbol{-}$}};
\node at (1.0,-1.2) {\Large{$\boldsymbol{-}$}};
\draw[<-, line width=1.5pt, blue] (-7,0)--(0,0); 
\node[blue] at (-3.5,0.3) {\Large{$rv$}};
\draw[blue] (-6.4,0) arc[start angle=0, end angle=90, radius=0.6];
\filldraw[blue] (-6.73,0.28) circle(0.04);
\filldraw[black] (0,0) circle(0.07);
\node[black] at (0.3,0.0) {\Large{$0$}};
\node at (-9.0,0.0) {\Large{Plate}};
\draw[->, line width=0.8pt, blue] (0,0)--(0.4,2.0); 
\draw[->, line width=0.8pt, blue] (0,0)--(-0.4,-2.0); 
\node[blue] at (0.5,0.95) {\large{$R_1$}};
\node[blue] at (-1.3,-0.95) {\large{$R_2=-R_1$}};
\end{tikzpicture}
\end{center}
\caption{\label{Fig:PlateWithHydrogen} A hydrogen molecule interacting with a perfectly conducting plate.}
\end{figure}
The Hamiltonian of the system consists of the Hamiltonian of the $\ce{H2}$ molecule and some more interaction terms with mirror charges, see e.g. \cite{ABH} for more details. 
In this setting the asymptotic behavior of the interaction energy $W(r,v)$ between the $\ce{H2}$ molecule and the plate has been estimated in \cite[Theorem 2.3]{ABH} assuming an energy condition (see \cite[Equation (2.28)]{ABH}), which we discuss and prove below. More precisely, it has been shown in \cite{ABH} that 
\begin{align}
W(r,v)= - \dfrac{C(v)}{r^3} + O\left( \dfrac{1}{r^4}\right) \qquad \qquad (r \to \infty)\,,   \label{PlateMoleculeInteraction} 
\end{align}
where the constant $C(v)>0$ depends on the orientation $v$ of the molecule and the form of its ground states, see \cite[Equation (2.36)]{ABH} for the precise definition of $C(v)$.
Equation \eqref{PlateMoleculeInteraction} states that the 
long-range behavior of the interaction  between the plate and the $\ce{H2}$ molecule is attractive and of the leading order $1/r^3$.
We now state our last main theorem, which is a slightly stronger form of the energy condition from \cite[Equation (2.28)]{ABH} in the special case of an $\ce{H2}$ molecule. 
\begin{theorem} \label{Thm:EnergyCondition}
Let $E(2,(1,1))$ and $E(1,(1,1))$ be defined by Eq. \eqref{DefEabs}. Then,
\begin{align*}
E(2,(1,1)) < E(1,(1,1)) + \dfrac{\Ea(1,1)}{16}\,. 
\end{align*}  
\end{theorem}

\begin{remark}
    Note that $E(2,(1,1))$ corresponds to the ground-state energy of the hydrogen molecule $\ce{H2}$ and $ E(1,(1,1))$ corresponds to the ground-state energy of the hydrogen molecule ion $\ce{H2+}$. The term $\Ea(1,1)/16$ is the minimal energy of a free electron in a half-space with a perfectly conducting boundary, see \cite[Equation (3.2)]{ABH} and its proof. As a consequence,
the physical meaning of Theorem \ref{Thm:EnergyCondition} is that when the plate and the $H_2$ molecule are far from each other, then it is more favorable for the electrons to stay close to the nuclei rather than to move to the plate. 
Although it is conjectured in \cite{ABH} that this condition holds in general, it was only analytically verified in the special cases of a hydrogen atom ($N=Z=1$) and a helium atom ($N=Z=2$). In particular, it was not known for any molecule.
 Note that Theorem \ref{Thm:EnergyCondition} is even stronger than what is needed in \cite{ABH}, as in \cite{ABH} it was only assumed that $E(2,(1,1),R) < E(1,(1,1),R) + \dfrac{\Ea(1,1)}{16}$ with $R$ being a minimizer of $E(2,(1,1),R)$.
\end{remark}
\begin{remark}
      As mentioned above, a major difficulty in the proof of Theorem \ref{Thm:EnergyCondition}, which is absent from the proof of Theorem \ref{theohydrogenhelium}, is that the binding lengths of the hydrogen molecule $\ce{H2}$ and the hydrogen molecule ion $\ce{H2+}$ are not known. One of the  ingredients in overcoming this difficulty is Estimate \eqref{Rbestimatediatomicexplicit}, which in the special case of $\ce{H2+}$ coincides with Estimate \eqref{BenguriaEestimateDiatomic}.
\end{remark}

%\noindent\textit{Remark.} 
 
The outline of this paper is as follows: Section \ref{secproofvdW} is devoted to the proof of Theorem \ref{theohydrogenhelium}. The proof of Theorem \ref{theoRbbound2} is given in Section \ref{secRbbound}. In Section \ref{Hydrogenplate} we  prove 
Theorem  \ref{Thm:EnergyCondition}.

\section{Proof of Theorem \ref{theohydrogenhelium}} \label{secproofvdW}
We divide the proof of Theorem \ref{theohydrogenhelium} into several steps, starting by recalling the proof for the pure hydrogen gas. 
Note that in case of the pure hydrogen gas, where $N_i=1$ for $i=1,\ldots,K$, Condition (E)  follows from the energy estimate (see also \cite[Proposition A.1]{AnapolitanosSigal2013})
\begin{align}
N\Ea(1,1) < \Ea(N,1) \qquad \qquad (N \geq 2)\,. \label{HydrogenBoundAS}
\end{align}
Instead of just proving Estimate \eqref{HydrogenBoundAS}, we adopt the proof from \cite[Proposition A.1]{AnapolitanosSigal2013} to prove a more general estimate that will also be helpful in the proofs of Theorem \ref{theohydrogenhelium} and Proposition \ref{propHelium} below.

\begin{lemma}\label{LemmaEN}
Let $Z>0$, $K \geq 2$ and $N_i \in \N$ for $i=1,\ldots,K$. Then,
\begin{align*}
\sum_{i=1}^{K}\Ea(N_i,Z)  < \Ea\left(\sum_{i=1}^{K} N_i,Z\right)\,.
\end{align*}
\end{lemma}

\noindent\textit{Remark.} A special case of Lemma \ref{LemmaEN} is the estimate
\begin{align*}
\Ea(N_1,Z)-\Ea(N_1+1,Z)<-\Ea(1,Z) \qquad \qquad (Z>0)\,,
\end{align*}
which shows that the last ionization energy of an atom is larger than any other ionization energy or electron affinity of the atom (see also \cite{Briet}).

\begin{proof}
Note that it suffices to prove the desired estimate for $K=2$, that is
\begin{align}
\Ea(N_1,Z) + \Ea(N_2,Z)  < \Ea\left(N_1+N_2,Z\right) \label{Estimaten=2}
\end{align}
for all $Z>0$ and $N_1, N_2 \in \N$, since the general case then follows by induction.
To prove Estimate \eqref{Estimaten=2}, we first assume that $\Ha(N_1+N_2,Z)$ has a normalized ground state, which we denote by $\psi_0$. Then,
\begin{align*}
\Ea(N_1+N_2,Z) =& \Braket{\psi_0 | \Ha(N_1+N_2,Z) \psi_0}\\
 =& \Braket{\psi_0 | \sum_{i=1}^{N_1}\left( -\Delta_{x_i} - \frac{Z}{\left|x_i\right|}\right) + \sum_{1 \leq i< j \leq N_1} \frac{1}{\left|x_i-x_j\right|} | \psi_0} \\
&+ \Braket{\psi_0 | \sum_{i=N_1+1}^{N_1+N_2}\left( -\Delta_{x_i} - \frac{Z}{\left|x_i\right|}\right) + \sum_{N_1+1 \leq i< j \leq N_1+N_2} \frac{1}{\left|x_i-x_j\right|} | \psi_0}
\\&+ \Braket{\psi_0 | \sum_{1 \leq i \leq N_1 <j \leq N_1+N_2} \frac{1}{\left|x_i-x_j\right|} | \psi_0}
\\>& \Ea(N_1,Z) + \Ea(N_2,Z)\,,
\end{align*}
where we used the positivity of the interaction term among the first $N_1$ electrons and the remaining $N_2$ electrons to obtain the strict inequality.

It remains to consider the case where $\Ha(N_1+N_2,Z)$ does not have a normalized ground state. Note that the HVZ theorem (see e.g. \cite[Theorem 11.2]{Teschl}) then implies that 
\begin{align}
\Ea(N_1+N_2,Z)=\Ea(N_1+N_2-1,Z)\,, \label{HVZidentity}
\end{align}
so Estimate \eqref{Estimaten=2} now follows by induction over $N_1+N_2 \geq 2$: For $N_1+N_2=2$, it follows from Eq. \eqref{HVZidentity} that Estimate \eqref{Estimaten=2} is equivalent to $\Ea(1,Z)<0$, which is obviously true due to the explicit formula $\Ea(1,Z)=-Z^2/4$. For $N_1+N_2\geq 3$, we may assume that $N_1 \geq 2$ without loss of generality. Estimate \eqref{Estimaten=2} then follows from 
\begin{align*}
\Ea(N_1+N_2,Z)= \Ea(N_1+N_2-1,Z) &> \Ea(N_1-1,Z) + \Ea(N_2,Z) \\
&\geq \Ea(N_1,Z) + \Ea(N_2,Z)\,,
\end{align*}
where the induction hypothesis was used for the first inequality and the second inequality used the fact that $\Ea(N,Z)$ is monotonically decreasing in the number of electrons $N \in \N$.
\end{proof}

In the next step, we consider the pure helium gas, which is also of physical relevance because in nature helium is monoatomic, so this can be viewed as a model for a real quantum gas. Instead of just proving Condition (E), we prove that even Condition (E') holds in this case:

\begin{proposition} \label{propHelium}
For all $m,n \in \N_0$ with $n \leq 1$, we have
\begin{align*}
\Ea(2+m,2)+\Ea(2-n,2) < \Ea(3+m,2) + \Ea(1-n,2) \qquad \qquad (\textup{E'}\,)\,.
\end{align*}
\end{proposition}

\noindent\textit{Remark.} Proposition \ref{propHelium} does not show that
\begin{align*}
\Ea(3,2)-\Ea(4,2) < \Ea(2,2)-\Ea(3,2)\,.
\end{align*}
The meaning of the last inequality is that the fourth last ionization energy is smaller than the third last, which is still an open conjecture (see also \cite{Briet}).\newline

To prove Proposition \ref{propHelium}, we need strong enough upper and lower bounds for the ground-state energies $\Ea(N,2)$ with $N \leq 4$. For the upper bounds, we use the standard textbook estimate (see e.g. \cite[Applications 4.3.21]{Thirring})
\begin{align}
\Ea(2,Z) \leq -\frac{1}{2}\left(Z-\frac{5}{16} \right)^2 \qquad \qquad \left(Z \geq \frac{5}{16}\right)\,, \label{EUpper}
\end{align}
which in particular implies that
\begin{align}
\Ea(2,2) \leq -\frac{729}{512}\,. \label{EUpperHelium}
\end{align}
For the convenience of the reader, we give a short proof of Estimate \eqref{EUpper}  in Lemma \ref{lmUpperproof} in the appendix. For the lower bounds, we use two different approaches. For the ground-state energy of $\textup{H}^{-}$, we use the lower bound from \cite[Eq. (4)]{Briet} resulting from the Bazley-Fox method, which yields the estimate
\begin{align}
\Ea(2,1) \geq -0.2887\,. \label{E21lower}
\end{align}
For the lower bounds for $\Ea(3,2)$ and $\Ea(4,2)$, we apply the recursive method that has been recently developed in \cite[Theorem 2.1]{LowerBoundsAtoms}. A simplified version of this method that is sufficient for our purposes is derived in Corollary \ref{korLower} below. A crucial ingredient for this method are upper bounds for the number of bound electrons 
\begin{align}
N_c(Z):= \max\left\{N \in \N \,: \, \Ea(N,Z) < \Ea(N-1,Z) \right\}. \label{DefNc}
\end{align}
The interpretation of Eq. \eqref{DefNc} is that $N_c(Z)$ is the maximum number of electrons that can be attached to the nucleus with a decrease in the ground-state energy\footnote{
In contrast to the strict inequality $E(N,Z)< E(N-1,Z)$, the inequality $E(N,Z)\leq E(N-1,Z)$ is true for all $N \in \N$ and $Z>0$.}.
Although the method from \cite[Theorem 2.1]{LowerBoundsAtoms} allows for  general upper bounds $N_c(Z) < f(Z)$, Lieb's bound  \cite{LiebIonization}, namely
\begin{align}
N_c(Z) < 2Z+1 \qquad \qquad (Z>0) \label{Liebbound},
\end{align}
 is sufficient for our purposes. 
From Eqs. \eqref{DefNc} and \eqref{Liebbound} we obtain that
$\Ea(N,Z)=\Ea(N-1,Z)$ for all $0<Z\leq (N-1)/2$.
Therefore, we obtain from \cite[Proposition 4.5]{LowerBoundsAtoms} the following simplified version of \cite[Theorem 2.1]{LowerBoundsAtoms}, where $\sigma_N$ was defined in Eq. \eqref{DefSigmaN}: 

\begin{kor} \label{korLower}
For all $N \geq 2$ and $Z>0$, we have
\begin{align*}
\sqrt{|\Ea(N,Z)|} \leq \begin{cases}
\sqrt{|\Ea(N-1,Z)|} & \qquad \textup{if} \; 0 < Z \leq \frac{N-1}{2}\,, \\
\sqrt{\left| \Ea\left(N-1,\frac{N-1}{2}\right) \right|} 
+ \sqrt{\sigma_{N}} \left(Z-\frac{N-1}{2}\right)  & \qquad \textup{if}\; Z>\frac{N-1}{2}\,.
\end{cases}
\end{align*}
\end{kor}

To derive lower bounds for $\Ea(3,2)$ and $\Ea(4,2)$, we 
apply Corollary \ref{korLower} recursively, starting with the initial estimate $ |\Ea(2,1)| \leq 0.2887$ from 
\eqref{E21lower}.  This results in the lower bounds 
\begin{align}
\Ea(3,2) &\geq -\left(\sqrt{|\Ea(2,1)|}+\sqrt{\sigma_3} \right)^2 \geq -1.6572\,, \label{E32lower} \\
\Ea(4,2) &\geq -\left(\sqrt{|\Ea(2,1)|}+\dfrac{\sqrt{\sigma_3} + \sqrt{\sigma_4}}{2} \right)^2  \geq -1.7098\,. \label{E42lower}
\end{align}

\begin{remark}\label{BrietImprovement}
    We note that Estimate \eqref{E32lower} improves the corresponding lower bound obtained by \cite[Eqs. (3) and (4)]{Briet} by about 5\%. This means that our proof improves the gap between the first ionization energy and the first electron affinity of the helium atom. 
\end{remark}
\begin{proof}[Proof of Proposition \ref{propHelium}]
First, we show that we may assume that $m \leq 1$ without restriction. To this end, we first note that for fixed $Z \in \N$ Lieb's bound \eqref{Liebbound} implies that increasing the number of electrons above $2Z$ does not further decrease the ground-state energy, i.e.
\begin{align}\label{Liebcorollary}
\Ea(N,Z)=\Ea(2Z,Z)  \qquad \qquad (N \geq  2Z)\,. 
\end{align}
In particular, it follows that $\Ea(2+m,2)= \Ea(3+m,2)$ for all $m \geq 2$. Since we also have that $\Ea(2-n,2) < \Ea(1-n,2)$, we see that Condition (E') is trivial for $m \geq 2$, so we can indeed assume that $m \leq 1$.  
This reduces the proof of Condition (E') to only two cases,
which we can now treat straightforwardly with the help of the above preparations: 
\medskip\newline\underline{Case 1: $n = 0$, $m\in \left\{0,1\right\}$}\newline
In this case, the desired estimate
\begin{align*}
\Ea(2+m,2) +  \Ea(2,2) < \Ea(3+m,2)+ \Ea(1,2)
\end{align*}
follows from the identity $\Ea(1,2)=-1$, the lower bound
\begin{align*}
\Ea(3+m,2)& \geq \Ea(4,2) \geq -1.7098\,,
\end{align*}
where the last estimate is the lower bound for $\Ea(4,2)$ from Eq. \eqref{E42lower}, and the upper bound $\Ea(2+m,2)+ \Ea(2,2) \leq 2\Ea(2,2) \leq  -729/256$, where the last estimate results from Eq. \eqref{EUpperHelium}.
\medskip\newline\underline{Case 2: $n=1$, $m \in \left\{0,1\right\}$}\newline
In this case, the desired estimate
\begin{align*}
\Ea(2+m,2) +  \Ea(1,2) < \Ea(3+m,2)
\end{align*}
is just a special case of Lemma \ref{LemmaEN}.
\end{proof}

\begin{proof}[Proof of Theorem \ref{theohydrogenhelium}]
We have to show that there is no distribution of the electrons that satisfies all the hypotheses of Theorem \ref{theohydrogenhelium}, but
\begin{align}
\sum_{i=1}^K \Ea(N_i,N_i) \geq  \sum_{i=1}^K \Ea(N_i+n_i,N_i)\,. \label{Eviolate}
\end{align}
Using again \eqref{Liebcorollary}, we can argue
 as in the proof of Proposition \ref{propHelium} that we may assume that $n_i \leq N_i$ for $i=1,...,K$. Furthermore, we can also assume that $n_i \neq 0$ for $i=1,...,K$ because otherwise the contributions of the atoms can simply be subtracted on both sides. After these preparations, it is now straightforward to show that the system after the electron transfer (corresponding to the right-hand side of Eq. \eqref{Eviolate}) can be divided into subsystems so that each subsystem has total charge zero and it is composed of at most three ions. Therefore, it suffices to prove the analog of Condition (E) for all these subsystems. More precisely, we have to consider the following four cases:
\medskip\newline
 \underline{Case 1: 2+, 2-}\medskip\newline
A doubly negatively charged ion and a doubly positively charged ion correspond to a subsystem composed of $\text{He}^{2+}$ and $\text{He}^{2-}$.
For such a subsystem, we have even verified the stronger Condition (E') in Proposition \ref{propHelium}.
\medskip\newline
\underline{Case 2: +, -}\medskip\newline
The two subsystems $(\text{H}^{+}, \text{H}^{-})$ and
$(\text{He}^{+}, \text{He}^{-})$ and the corresponding
energy estimates
\begin{align*}
2\Ea(1,1) &< \Ea(2,1)\,, \\ 
2\Ea(2,2) &< \Ea(1,2) + \Ea(3,2)
\end{align*}
have already been verified in Lemma \ref{LemmaEN} and Proposition \ref{propHelium}, respectively. It remains to consider all subsystems, which contain both hydrogen and helium ions. These are the two subsystems $(\text{He}^{+}, \text{H}^{-})$ and $(\text{H}^{+}, \text{He}^{-})$ resulting from the (hypothetical) electron transfers illustrated in Figure \ref{Fig:+-} below. 

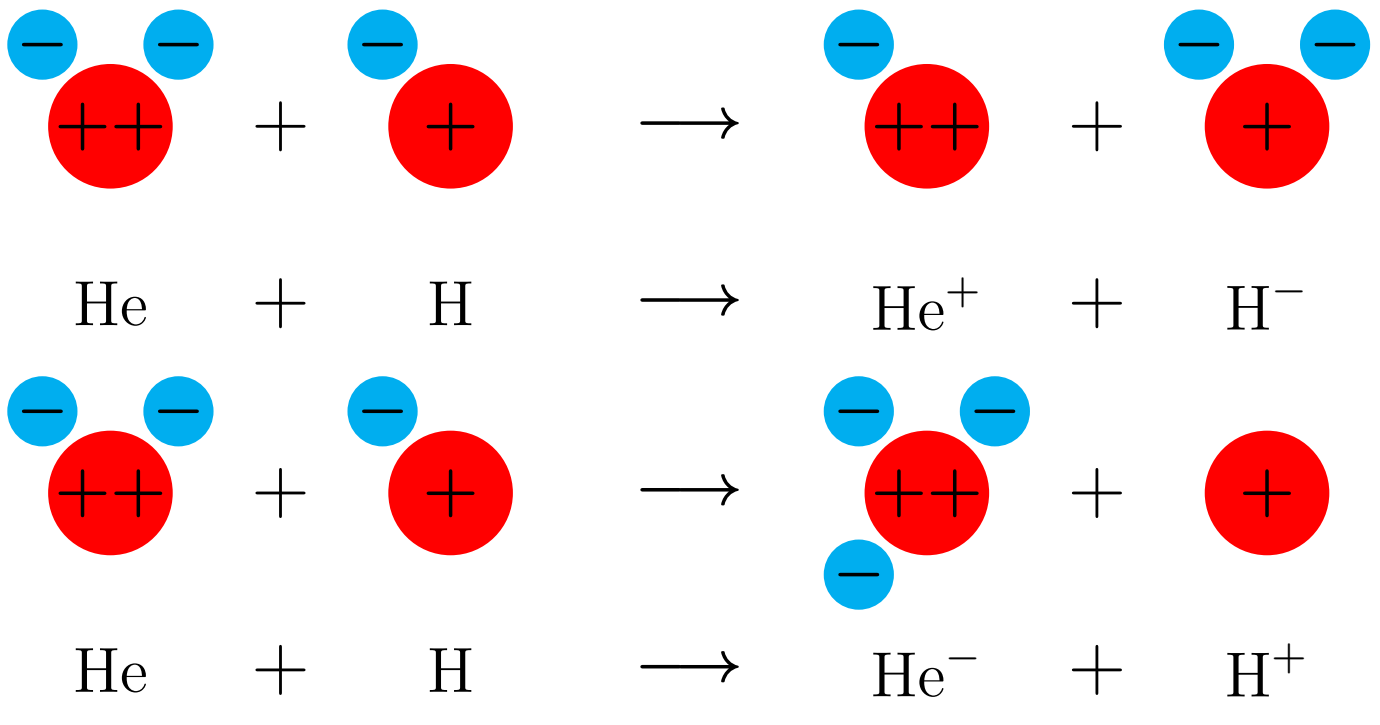
\begin{figure}[H]
\begin{tikzpicture}[scale=0.9]
\filldraw[red] (0.0,0.0) circle(4.5mm);
\node at (0.0,-0.0) {\large{$\boldsymbol{++}$}};
\filldraw[cyan] (0.5,0.6) circle(2.5mm);
\node at (0.5,0.6) {$\boldsymbol{-}$}; 
\filldraw[cyan] (-0.5,0.6) circle(2.5mm);
\node at (-0.5,0.6) {$\boldsymbol{-}$}; 
\node at (0.0,-1.3) {\large {He}}; 

\node at (1.25,0.0) {\Large {$+$}}; 
\node at (1.25,-1.3) {\Large {$+$}}; 

\filldraw[red] (2.5,0.0) circle(4.5mm);
\node at (2.5,-0.0) {\large{$\boldsymbol{+}$}};
\filldraw[cyan] (2.0,0.6) circle(2.5mm);
\node at (2.0,0.6) {$\boldsymbol{-}$}; 
\node at (2.5,-1.3) {\large {H}}; 

\node at (4.25,0.0) {$\boldsymbol{\longrightarrow}$}; 
\node at (4.25,-1.3) {$\boldsymbol{\longrightarrow}$}; 

\node at (7.25,0.0) {\Large {$+$}}; 
\node at (7.25,-1.3) {\Large {$+$}}; 

\filldraw[red] (6.0,0.0) circle(4.5mm);
\node at (6.0,-0.0) {\large{$\boldsymbol{++}$}};
\filldraw[cyan] (5.5,0.6) circle(2.5mm);
\node at (5.5,0.6) {$\boldsymbol{-}$};
\node at (6.0,-1.3) {\large {$\text{He}^{+}$}}; 

\filldraw[red] (8.5,0.0) circle(4.5mm);
\node at (8.5,-0.0) {\large{$\boldsymbol{+}$}};
\filldraw[cyan] (8.0,0.6) circle(2.5mm);
\node at (8.0,0.6) {$\boldsymbol{-}$};
\filldraw[cyan] (9.0,0.6) circle(2.5mm);
\node at (9.0,0.6) {$\boldsymbol{-}$}; 
\node at (8.5,-1.3) {\large {$\text{H}^{-}$}};
\end{tikzpicture}
\vspace*{1mm}\newline
\begin{tikzpicture}[scale=0.9]
\filldraw[red] (0.0,0.0) circle(4.5mm);
\node at (0.0,-0.0) {\large{$\boldsymbol{++}$}};
\filldraw[cyan] (0.5,0.6) circle(2.5mm);
\node at (0.5,0.6) {$\boldsymbol{-}$}; 
\filldraw[cyan] (-0.5,0.6) circle(2.5mm);
\node at (-0.5,0.6) {$\boldsymbol{-}$}; 
\node at (0.0,-1.3) {\large {He}}; 

\node at (1.25,0.0) {\Large {$+$}}; 
\node at (1.25,-1.3) {\Large {$+$}}; 

\filldraw[red] (2.5,0.0) circle(4.5mm);
\node at (2.5,-0.0) {\large{$\boldsymbol{+}$}};
\filldraw[cyan] (2.0,0.6) circle(2.5mm);
\node at (2.0,0.6) {$\boldsymbol{-}$}; 
\node at (2.5,-1.3) {\large {H}}; 

\node at (4.25,0.0) {$\boldsymbol{\longrightarrow}$}; 
\node at (4.25,-1.3) {$\boldsymbol{\longrightarrow}$}; 

\node at (7.25,0.0) {\Large {$+$}}; 
\node at (7.25,-1.3) {\Large {$+$}}; 

\filldraw[red] (6.0,0.0) circle(4.5mm);
\node at (6.0,-0.0) {\large{$\boldsymbol{++}$}};
\filldraw[cyan] (5.5,0.6) circle(2.5mm);
\node at (5.5,0.6) {$\boldsymbol{-}$};
\filldraw[cyan] (6.5,0.6) circle(2.5mm);
\node at (6.5,0.6) {$\boldsymbol{-}$};
\filldraw[cyan] (5.5,-0.6) circle(2.5mm);
\node at (5.5,-0.6) {$\boldsymbol{-}$};
\node at (6.0,-1.3) {\large {$\text{He}^{-}$}}; 

\filldraw[red] (8.5,0.0) circle(4.5mm);
\node at (8.5,-0.0) {\large{$\boldsymbol{+}$}};
\node at (8.5,-1.3) {\large {$\text{H}^{+}$}};
\end{tikzpicture}
\caption{\label{Fig:+-} The (hypothetical) electron transfers leading to the subsystems $(\text{He}^{+}, \text{H}^{-})$ and $(\text{H}^{+}, \text{He}^{-})$}
\end{figure}

\noindent For the subsystem  $(\text{He}^{+}, \text{H}^{-})$ Condition (E)  reads
\begin{align*}
\Ea(2,2) + \Ea(1,1) < \Ea(1,2) + \Ea(2,1)\,, 
\end{align*}
and for the subsystem $(\text{H}^{+}, \text{He}^{-})$ Condition (E)  reads
\begin{align*}
\Ea(2,2) + \Ea(1,1) < \Ea(3,2)\,.
\end{align*}
\noindent Both of these estimates can be obtained from the estimates that we have already used in the proof of Proposition \ref{propHelium}: To estimate the left-hand sides, we use the identity $\Ea(1,1)=-1/4$ and the upper bound \eqref{EUpperHelium} for $\Ea(2,2)$. To estimate the right-hand sides, we use the identity $\Ea(1,2)=-1$ and the lower bounds for $\Ea(2,1)$ and $\Ea(3,2)$ from Estimates \eqref{E21lower} and \eqref{E32lower}, respectively.\medskip\newline 
\underline{Case 3: 2+, -, -}\medskip\newline
The subsystem $(\text{He}^{2+}, \text{He}^{-},\text{He}^{-})$ only involves helium nuclei, so it has already been treated in Proposition \ref{propHelium}. 
Consequently, we only have to consider the mixed  
subsystems  $(\text{He}^{2+}, \text{H}^{-},\text{H}^{-})$
and $(\text{He}^{2+}, \text{He}^{-},\text{H}^{-})$ resulting from the (hypothetical) electron transfers illustrated in Figure \ref{Fig:2+--} below. 

\begin{figure}[H] \label{Fig:4}
\begin{tikzpicture}[scale=0.9]
\filldraw[red] (0.0,0.0) circle(4.5mm);
\node at (0.0,-0.0) {\large{$\boldsymbol{++}$}};
\filldraw[cyan] (0.5,0.6) circle(2.5mm);
\node at (0.5,0.6) {$\boldsymbol{-}$}; 
\filldraw[cyan] (-0.5,0.6) circle(2.5mm);
\node at (-0.5,0.6) {$\boldsymbol{-}$}; 
\node at (0.0,-1.3) {\large {He}}; 

\node at (1.25,0.0) {\Large {$+$}}; 
\node at (1.25,-1.3) {\Large {$+$}}; 

\filldraw[red] (2.5,0.0) circle(4.5mm);
\node at (2.5,-0.0) {\large{$\boldsymbol{+}$}};
\filldraw[cyan] (2.0,0.6) circle(2.5mm);
\node at (2.0,0.6) {$\boldsymbol{-}$}; 
\node at (2.5,-1.3) {\large {H}}; 

\node at (3.75,0.0) {\Large {$+$}}; 
\node at (3.75,-1.3) {\Large {$+$}}; 

\filldraw[red] (5.0,0.0) circle(4.5mm);
\node at (5.0,-0.0) {\large{$\boldsymbol{+}$}};
\filldraw[cyan] (4.5,0.6) circle(2.5mm);
\node at (4.5,0.6) {$\boldsymbol{-}$}; 
\node at (5.0,-1.3) {\large {H}}; 

\node at (6.75,0.0) {$\boldsymbol{\longrightarrow}$}; 
\node at (6.75,-1.3) {$\boldsymbol{\longrightarrow}$}; 

\filldraw[red] (8.5,0.0) circle(4.5mm);
\node at (8.5,-0.0) {\large{$\boldsymbol{++}$}};
\node at (8.5,-1.3) {\large {$\text{He}^{2+}$}}; 

\node at (9.75,0.0) {\Large {$+$}}; 
\node at (9.75,-1.3) {\Large {$+$}}; 

\filldraw[red] (11.0,0.0) circle(4.5mm);
\node at (11.0,-0.0) {\large{$\boldsymbol{+}$}};
\filldraw[cyan] (11.5,0.6) circle(2.5mm);
\node at (11.5,0.6) {$\boldsymbol{-}$};
\filldraw[cyan] (10.5,0.6) circle(2.5mm);
\node at (10.5,0.6) {$\boldsymbol{-}$};
\node at (11.0,-1.3) {\large {$\text{H}^{-}$}}; 

\node at (12.25,0.0) {\Large {$+$}}; 
\node at (12.25,-1.3) {\Large {$+$}}; 

\filldraw[red] (13.5,0.0) circle(4.5mm);
\node at (13.5,-0.0) {\large{$\boldsymbol{+}$}};
\filldraw[cyan] (14.0,0.6) circle(2.5mm);
\node at (14.0,0.6) {$\boldsymbol{-}$};
\filldraw[cyan] (13.0,0.6) circle(2.5mm);
\node at (13.0,0.6) {$\boldsymbol{-}$};
\node at (13.5,-1.3) {\large {$\text{H}^{-}$}};
\end{tikzpicture}
\vspace*{1mm}\newline

\begin{tikzpicture}[scale=0.9]
\filldraw[red] (0.0,0.0) circle(4.5mm);
\node at (0.0,-0.0) {\large{$\boldsymbol{++}$}};
\filldraw[cyan] (0.5,0.6) circle(2.5mm);
\node at (0.5,0.6) {$\boldsymbol{-}$}; 
\filldraw[cyan] (-0.5,0.6) circle(2.5mm);
\node at (-0.5,0.6) {$\boldsymbol{-}$}; 
\node at (0.0,-1.3) {\large {He}}; 

\node at (1.25,0.0) {\Large {$+$}}; 
\node at (1.25,-1.3) {\Large {$+$}}; 

\filldraw[red] (2.5,0.0) circle(4.5mm);
\node at (2.5,-0.0) {\large{$\boldsymbol{++}$}};
\filldraw[cyan] (3.0,0.6) circle(2.5mm);
\node at (3.0,0.6) {$\boldsymbol{-}$}; 
\filldraw[cyan] (2.0,0.6) circle(2.5mm);
\node at (2.0,0.6) {$\boldsymbol{-}$}; 
\node at (2.5,-1.3) {\large {He}}; 

\node at (3.75,0.0) {\Large {$+$}}; 
\node at (3.75,-1.3) {\Large {$+$}}; 

\filldraw[red] (5.0,0.0) circle(4.5mm);
\node at (5.0,-0.0) {\large{$\boldsymbol{+}$}};
\filldraw[cyan] (4.5,0.6) circle(2.5mm);
\node at (4.5,0.6) {$\boldsymbol{-}$}; 
\node at (5.0,-1.3) {\large {H}}; 

\node at (6.75,0.0) {$\boldsymbol{\longrightarrow}$}; 
\node at (6.75,-1.3) {$\boldsymbol{\longrightarrow}$}; 

\filldraw[red] (8.5,0.0) circle(4.5mm);
\node at (8.5,-0.0) {\large{$\boldsymbol{++}$}};
\node at (8.5,-1.3) {\large {$\text{He}^{2+}$}}; 

\node at (9.75,0.0) {\Large {$+$}}; 
\node at (9.75,-1.3) {\Large {$+$}}; 

\filldraw[red] (11.0,0.0) circle(4.5mm);
\node at (11.0,-0.0) {\large{$\boldsymbol{++}$}};
\filldraw[cyan] (11.5,0.6) circle(2.5mm);
\node at (11.5,0.6) {$\boldsymbol{-}$};
\filldraw[cyan] (10.5,0.6) circle(2.5mm);
\node at (10.5,0.6) {$\boldsymbol{-}$};
\filldraw[cyan] (10.5,-0.6) circle(2.5mm);
\node at (10.5,-0.6) {$\boldsymbol{-}$};
\node at (11.0,-1.3) {\large {$\text{He}^{-}$}}; 

\node at (12.25,0.0) {\Large {$+$}}; 
\node at (12.25,-1.3) {\Large {$+$}}; 

\filldraw[red] (13.5,0.0) circle(4.5mm);
\node at (13.5,-0.0) {\large{$\boldsymbol{+}$}};
\filldraw[cyan] (14.0,0.6) circle(2.5mm);
\node at (14.0,0.6) {$\boldsymbol{-}$};
\filldraw[cyan] (13.0,0.6) circle(2.5mm);
\node at (13.0,0.6) {$\boldsymbol{-}$};
\node at (13.5,-1.3) {\large {$\text{H}^{-}$}};
\end{tikzpicture}
\caption{\label{Fig:2+--} The (hypothetical) electron transfers leading to the subsystems $(\text{He}^{2+}, \text{H}^{-},\text{H}^{-})$
and $(\text{He}^{2+}, \text{He}^{-},\text{H}^{-})$}
\end{figure}
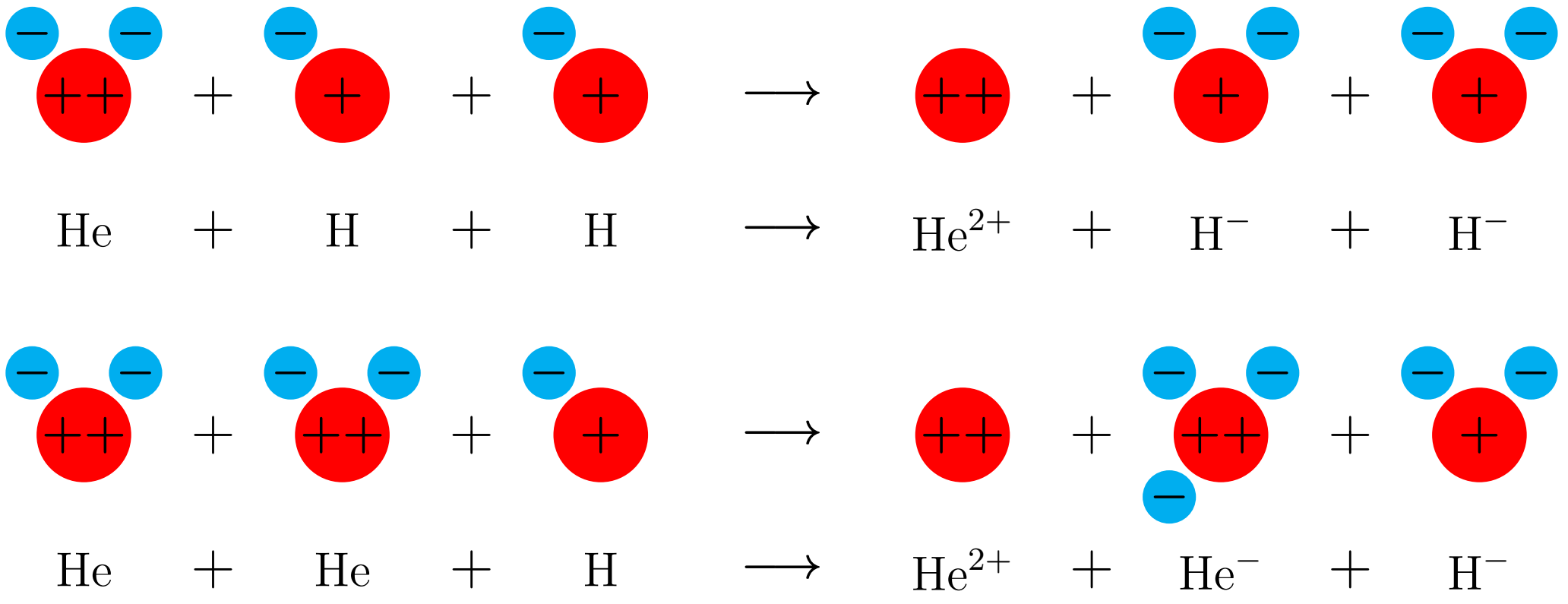

\noindent These subsystems correspond to the energy estimates 
\begin{align*}
\Ea(2,2) + 2\Ea(1,1) &< 2\Ea(2,1)\,, \\
2\Ea(2,2) + \Ea(1,1) &< \Ea(3,2) + \Ea(2,1)\,.
\end{align*}
Both estimates follow from the bounds that we have already used in Case 2, namely the identity $\Ea(1,1)=-1/4$, 
the upper bound \eqref{EUpperHelium} for $\Ea(2,2)$, 
and the lower bounds for $\Ea(2,1)$ and $\Ea(3,2)$ from Estimates \eqref{E21lower} and \eqref{E32lower}, respectively.\medskip\newline
\underline{Case 4: +, +, 2-}\medskip\newline
The subsystem $(\text{He}^{+}, \text{He}^{+},\text{He}^{2-})$ has already been treated in Proposition \ref{propHelium}. Therefore, it only remains to consider the subsystems  
$(\text{H}^{+}, \text{H}^{+},\text{He}^{2-})$ and
$(\text{He}^{+}, \text{H}^{+},\text{He}^{2-})$
 resulting from the (hypothetical) electron transfers illustrated in Figure \ref{Fig:++2-} below. 
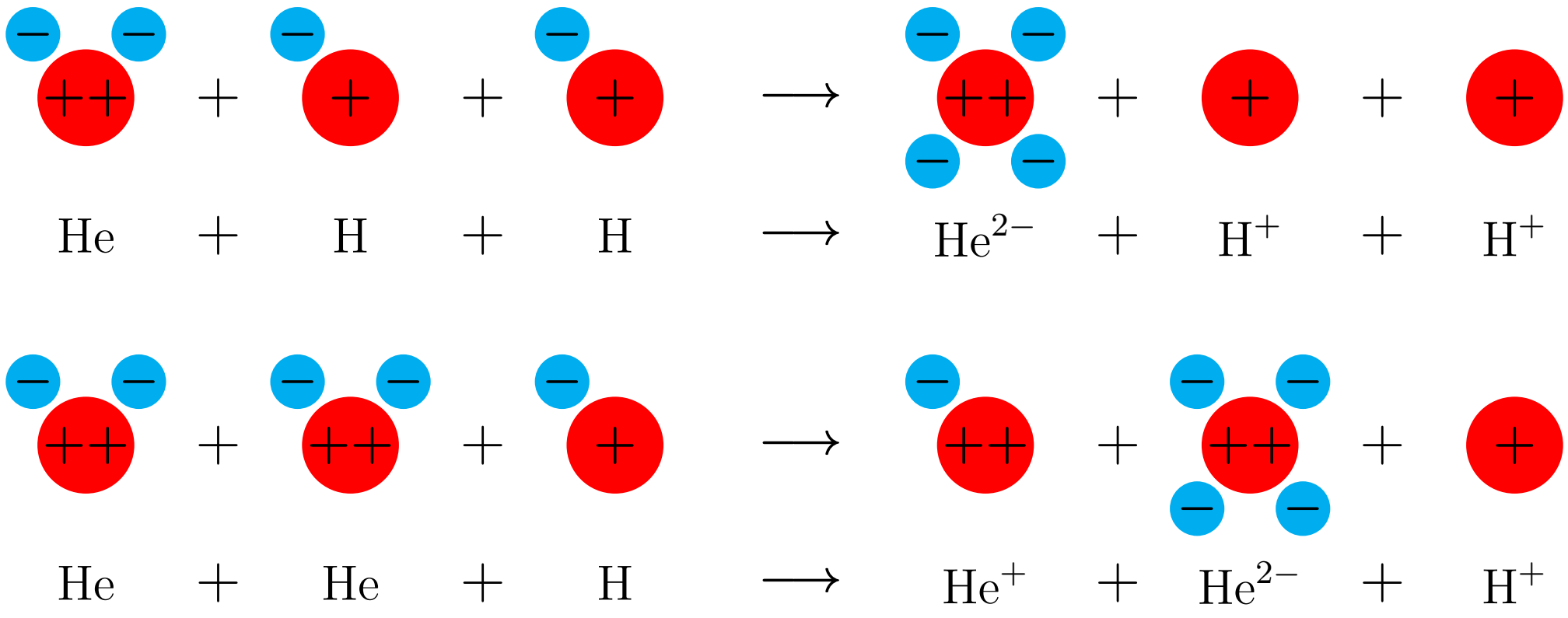
\begin{figure}[H]
\begin{tikzpicture}[scale=0.9]
\filldraw[red] (0.0,0.0) circle(4.5mm);
\node at (0.0,-0.0) {\large{$\boldsymbol{++}$}};
\filldraw[cyan] (0.5,0.6) circle(2.5mm);
\node at (0.5,0.6) {$\boldsymbol{-}$}; 
\filldraw[cyan] (-0.5,0.6) circle(2.5mm);
\node at (-0.5,0.6) {$\boldsymbol{-}$}; 
\node at (0.0,-1.3) {\large {He}}; 

\node at (1.25,0.0) {\Large {$+$}}; 
\node at (1.25,-1.3) {\Large {$+$}}; 

\filldraw[red] (2.5,0.0) circle(4.5mm);
\node at (2.5,-0.0) {\large{$\boldsymbol{+}$}};
\filldraw[cyan] (2.0,0.6) circle(2.5mm);
\node at (2.0,0.6) {$\boldsymbol{-}$}; 
\node at (2.5,-1.3) {\large {H}}; 

\node at (3.75,0.0) {\Large {$+$}}; 
\node at (3.75,-1.3) {\Large {$+$}}; 

\filldraw[red] (5.0,0.0) circle(4.5mm);
\node at (5.0,-0.0) {\large{$\boldsymbol{+}$}};
\filldraw[cyan] (4.5,0.6) circle(2.5mm);
\node at (4.5,0.6) {$\boldsymbol{-}$}; 
\node at (5.0,-1.3) {\large {H}}; 

\node at (6.75,0.0) {$\boldsymbol{\longrightarrow}$}; 
\node at (6.75,-1.3) {$\boldsymbol{\longrightarrow}$}; 

\filldraw[red] (8.5,0.0) circle(4.5mm);
\node at (8.5,-0.0) {\large{$\boldsymbol{++}$}};
\filldraw[cyan] (8.0,0.6) circle(2.5mm);
\node at (8.0,0.6) {$\boldsymbol{-}$};
\filldraw[cyan] (9.0,0.6) circle(2.5mm);
\node at (9.0,0.6) {$\boldsymbol{-}$};
\filldraw[cyan] (8.0,-0.6) circle(2.5mm);
\node at (8.0,-0.6) {$\boldsymbol{-}$};
\filldraw[cyan] (9.0,-0.6) circle(2.5mm);
\node at (9.0,-0.6) {$\boldsymbol{-}$};
\node at (8.5,-1.3) {\large {$\text{He}^{2-}$}}; 

\node at (9.75,0.0) {\Large {$+$}}; 
\node at (9.75,-1.3) {\Large {$+$}}; 

\filldraw[red] (11.0,0.0) circle(4.5mm);
\node at (11.0,-0.0) {\large{$\boldsymbol{+}$}};
\node at (11.0,-1.3) {\large {$\text{H}^{+}$}}; 

\node at (12.25,0.0) {\Large {$+$}}; 
\node at (12.25,-1.3) {\Large {$+$}}; 

\filldraw[red] (13.5,0.0) circle(4.5mm);
\node at (13.5,-0.0) {\large{$\boldsymbol{+}$}};
\node at (13.5,-1.3) {\large {$\text{H}^{+}$}};
\end{tikzpicture}
\vspace*{1mm}\newline

\begin{tikzpicture}[scale=0.9]
\filldraw[red] (0.0,0.0) circle(4.5mm);
\node at (0.0,-0.0) {\large{$\boldsymbol{++}$}};
\filldraw[cyan] (0.5,0.6) circle(2.5mm);
\node at (0.5,0.6) {$\boldsymbol{-}$}; 
\filldraw[cyan] (-0.5,0.6) circle(2.5mm);
\node at (-0.5,0.6) {$\boldsymbol{-}$}; 
\node at (0.0,-1.3) {\large {He}}; 

\node at (1.25,0.0) {\Large {$+$}}; 
\node at (1.25,-1.3) {\Large {$+$}}; 

\filldraw[red] (2.5,0.0) circle(4.5mm);
\node at (2.5,-0.0) {\large{$\boldsymbol{++}$}};
\filldraw[cyan] (3.0,0.6) circle(2.5mm);
\node at (3.0,0.6) {$\boldsymbol{-}$}; 
\filldraw[cyan] (2.0,0.6) circle(2.5mm);
\node at (2.0,0.6) {$\boldsymbol{-}$}; 
\node at (2.5,-1.3) {\large {He}}; 

\node at (3.75,0.0) {\Large {$+$}}; 
\node at (3.75,-1.3) {\Large {$+$}}; 

\filldraw[red] (5.0,0.0) circle(4.5mm);
\node at (5.0,-0.0) {\large{$\boldsymbol{+}$}};
\filldraw[cyan] (4.5,0.6) circle(2.5mm);
\node at (4.5,0.6) {$\boldsymbol{-}$}; 
\node at (5.0,-1.3) {\large {H}}; 

\node at (6.75,0.0) {$\boldsymbol{\longrightarrow}$}; 
\node at (6.75,-1.3) {$\boldsymbol{\longrightarrow}$}; 

\filldraw[red] (8.5,0.0) circle(4.5mm);
\node at (8.5,-0.0) {\large{$\boldsymbol{++}$}};
\filldraw[cyan] (8.0,0.6) circle(2.5mm);
\node at (8.0,0.6) {$\boldsymbol{-}$};
\node at (8.5,-1.3) {\large {$\text{He}^{+}$}}; 

\node at (9.75,0.0) {\Large {$+$}}; 
\node at (9.75,-1.3) {\Large {$+$}}; 

\filldraw[red] (11.0,0.0) circle(4.5mm);
\node at (11.0,-0.0) {\large{$\boldsymbol{++}$}};
\filldraw[cyan] (11.5,0.6) circle(2.5mm);
\node at (11.5,0.6) {$\boldsymbol{-}$};
\filldraw[cyan] (10.5,0.6) circle(2.5mm);
\node at (10.5,0.6) {$\boldsymbol{-}$};
\filldraw[cyan] (10.5,-0.6) circle(2.5mm);
\node at (10.5,-0.6) {$\boldsymbol{-}$};
\filldraw[cyan] (11.5,-0.6) circle(2.5mm);
\node at (11.5,-0.6) {$\boldsymbol{-}$};
\node at (11.0,-1.3) {\large {$\text{He}^{2-}$}}; 

\node at (12.25,0.0) {\Large {$+$}}; 
\node at (12.25,-1.3) {\Large {$+$}}; 

\filldraw[red] (13.5,0.0) circle(4.5mm);
\node at (13.5,-0.0) {\large{$\boldsymbol{+}$}};
\node at (13.5,-1.3) {\large {$\text{H}^{+}$}};
\end{tikzpicture}
\caption{\label{Fig:++2-} The (hypothetical) electron transfers leading to the subsystems $(\text{H}^{+}, \text{H}^{+},\text{He}^{2-})$ and
$(\text{He}^{+}, \text{H}^{+},\text{He}^{2-})$}
\end{figure}

\noindent These subsystems correspond to the energy estimates
\begin{align*}
\Ea(2,2) + 2\Ea(1,1) &< \Ea(4,2)\,,\\
2\Ea(2,2) + \Ea(1,1) &< \Ea(4,2) + \Ea(1,2)\,.
\end{align*}
Both  estimates follow from the identities $\Ea(1,1)=-1/4$, $\Ea(1,2)=-1$, the upper bound \eqref{EUpperHelium} for $\Ea(2,2)$ and the lower bound for $\Ea(4,2)$ from Estimate \eqref{E42lower}.
\end{proof}

\section{Proof of Theorem \ref{theoRbbound2}} \label{secRbbound}
The  proof  of Theorem \ref{theoRbbound2} is based on the united atom bound, which was also used in the proof of \cite[Theorem 1]{MoleculesBenguria}, and on the following lemma taken from  \cite[Theorem 4.6.15]{Thirring}.

\begin{lemma}[Thirring]\label{lmEUpper}
Let $K \geq 2$, $N \in \N$, $\underline{Z} \in (0,\infty)^K$ and $\underline{R} \in \R^{3K}$ be fixed. Then the function $\lambda \mapsto \lambda^2 E(N,\underline{Z},\lambda \underline{R})$ is concave in $\lambda>0$. Moreover, if $E(N,\underline{Z},\underline{R})$ has a local minimum at $\underline{R}=\underline{R}_0$, then the upper bound
\begin{align}
E(N,\underline{Z},\lambda\underline{R}_0) \leq  \frac{2 \lambda -1}{\lambda^2} E(N,\underline{Z},\underline{R}_0)  
 \label{TangentUpper}
\end{align}    
holds for all $\lambda>0$.
\end{lemma}

\begin{remark}
     Note that in \cite[Theorem 4.6.15]{Thirring} it is assumed that $\frac{d}{d\lambda}E(N,\underline{Z},\lambda \underline{R}_0)|_{\lambda=1}=0$, so $\underline{R}_0$ only has to be a critical point of the energy. However, for the argument to work, the existence of the derivative is assumed. Instead, we assume that   $\underline{R}_0$ is a local minimum and we can slightly adapt the argument in order to avoid assuming differentiability. 
\end{remark}

\begin{proof}[Proof of Lemma \ref{lmEUpper}]
For fixed $\psi  \in \mathcal{H}^2$, the scaling $x_i \rightarrow \lambda x_i$ for $i=1,\ldots,N$ leads to
\begin{align*}
\lambda^2 \Braket{\psi | H(N,\underline{Z}, \lambda \underline{R}) \psi} = \Braket{\psi_{\lambda} | \left( \,\sum_{i=1}^N -\Delta_{x_i} + \lambda V \right) \psi_{\lambda}}\,,
\end{align*}
where 
\begin{align}
\psi_{\lambda}(x_1,s_1,\ldots,x_N,s_N):= \lambda^{3N/2} \psi(\lambda x_1,s_1,\ldots, \lambda x_N,s_N) \label{Defphilambda}
\end{align}
and
\begin{align*}
V:=   \sum_{1 \leq i< j \leq N} \frac{1}{\left|x_i-x_j\right|}  + \sum_{1 \leq i < j \leq K} \frac{Z_i Z_j}{|R_i-R_j|}  -\sum_{i=1}^{N} \sum_{j=1}^{K} \frac{Z_j}{\left|x_i-R_j\right|}
\end{align*}
does not depend on $\lambda>0$. Hence, it follows that
\begin{align*}
\lambda^2 E(N,\underline{Z},\lambda \underline{R}) = \inf_{\stackrel{\psi \in \mathcal{H}^2}{\| \psi\|=1}}  \Braket{\psi | \left( \,\sum_{i=1}^N -\Delta_{x_i} + \lambda V \right) \psi}
\end{align*}
is concave as a function of $\lambda>0$,  because it is an infimum of affine functions in $\lambda$. It remains to prove Estimate \eqref{TangentUpper}, provided that $E(N,\underline{Z},\underline{R})$ has a local minimum at $\underline{R}=\underline{R}_0$. To this end, we first note that Estimate \eqref{TangentUpper} is equivalent to
\begin{align}
f(\lambda)-f(1) \leq 2(\lambda -1) f(1)\,,   \label{fBound}
\end{align}
where  $f(\lambda):= \lambda^2 E(N,\underline{Z},\lambda \underline{R}_0)$.
Due to the concavity of the function $f$, we find that
\begin{align}
f(\lambda)-f(1) \leq f'_{-}(1)(\lambda-1)\,,   \label{f1Bound}
\end{align}
where $f'_{-}$ denotes the left-hand derivative of $f$.
Observe now that
\begin{equation}\label{f2bound}
f'_{-}(1)= 2 E(N,\underline{Z},\underline{R}_0) + \frac{\partial}{\partial \lambda_{-}} \left( E(N,\underline{Z},\lambda\underline{R}_0)\right) \big\vert_{\lambda=1} \leq 2f(1)\,,
\end{equation}
where the inequality follows from $\frac{\partial}{\partial \lambda_{-}} \left( E(N,\underline{Z},\lambda\underline{R}_0)\right) \big\vert_{\lambda=1} \leq 0$ since $E(N,\underline{Z},\lambda \underline{R}_0)$ has a local minimum at $\lambda=1$. From Estimates \eqref{f1Bound} and \eqref{f2bound} we obtain the desired Estimate \eqref{fBound} for $\lambda \geq 1$. Estimate \eqref{fBound} for $0<\lambda < 1$ follows from an analog argument with $f'_{-}(1)$ replaced by $f'_{+}(1)$ (note that the equivalent of Estimate \eqref{f2bound} is then $f'_{+}(1) \geq 2f(1)$).
\end{proof}

\begin{proof}[Proof of Theorem \ref{theoRbbound2}]
Using  Estimate \eqref{TangentUpper} and the fact that $\lim_{\lambda \rightarrow 0} E(N,\underline{Z},\lambda\underline{R}_0)=\infty$, we find that $E(N,\underline{Z},\underline{R}_0)<0$. Note that a lower bound for $E(N,\underline{Z},\lambda\underline{R}_0)$ follows from the united atom bound (see e.g. \cite[Theorem 4.6.11]{Thirring}) 
\begin{align}
E(N,\underline{Z},\underline{R}) \geq \Ea(N,Z) + \sum_{1 \leq i< j \leq K} \frac{Z_i Z_j}{\left|R_{i}-R_{j}\right|}\,,  \label{UnitedAtom}
\end{align}
whose physical meaning is that if we exclude the nuclear repulsion, then the system has the lowest possible energy when all nuclei are at the same point.
Applying Estimate \eqref{UnitedAtom} for $\underline{R}=\underline{R}_0$, we find that 
$\Ea(N,Z)<E(N,\underline{Z},\underline{R}_0)$.
Moreover, using Estimates \eqref{TangentUpper} and \eqref{UnitedAtom} for $\underline{R}=\lambda \underline{R}_0$,  we obtain that
\begin{align*}
\Ea(N,Z) +   \sum_{1 \leq i< j \leq K} \frac{ Z_i Z_j}{\lambda\left|R_{0,i}-R_{0,j}\right|} \leq E(N,\underline{Z},\lambda \underline{R}_0) \leq \frac{2\lambda-1}{\lambda^2} E(N,\underline{Z}, \underline{R}_0) \,, \end{align*}
which implies that
\begin{align*}
  \sum_{1 \leq i< j \leq K} \frac{ Z_i Z_j}{\left|R_{0,i}-R_{0,j}\right|} \leq -\lambda \Ea(N,Z)  + \left(2-\frac{1}{\lambda}\right) E(N,\underline{Z}, \underline{R}_0) \,. \end{align*}
Optimizing the right-hand side with respect to $\lambda$ gives
\begin{align}
  \sum_{1 \leq i< j \leq K} \frac{ Z_i Z_j}{\left|R_{0,i}-R_{0,j}\right|} \leq 2 \sqrt{|E(N,\underline{Z}, \underline{R}_0)|} \left(\sqrt{|\Ea(N,Z)|} -  \sqrt{|E(N,\underline{Z},\underline{R}_0)|}\right) \,,
\end{align}
and after combining this with the estimate
\begin{align}
\sum_{1 \leq i< j \leq K} \frac{Z_i Z_j}{\left|R_{0,i}-R_{0,j}\right|} \geq \dfrac{1}{R_b} \sum_{1 \leq i< j \leq K} Z_i Z_j = \dfrac{Z^2- \sum_{i=1}^K (Z_i)^2}{2 R_b}\,,
\end{align}
we arrive at Estimate \eqref{Rbglobalestimate}.

\end{proof}

\section{Proof of Theorem \ref{Thm:EnergyCondition}}\label{Hydrogenplate}

\medskip 
An additional difficulty in proving Theorem \ref{Thm:EnergyCondition}, which does not arise in the proof of Theorem \ref{theohydrogenhelium}, stems from the fact that the minimizing configurations $R_2$ of $E(2,(1,1),R)$ and $R_1$ of $E(1,(1,1),R)$ are not known and are also expected to be not equal. To overcome this difficulty, we use the bound \eqref{Rbestimatediatomicexplicit} in the special case of  $\ce{H2+}$, from which it follows that the minimizer $R_1$ of $E(1,(1,1),R)$ satisfies the inequality $R_1 \geq 2$. 
Note that in this case our bound \eqref{Rbestimatediatomicexplicit} and the bound \eqref{BenguriaEestimateDiatomic} derived in \cite[THEOREM 2]{MoleculesBenguria} coincide. 
This helped us to derive 
a sufficiently good lower bound for the ground-state energy $E(1,(1,1))$ of the hydrogen molecule ion $\ce{H2+}$.

The Hamiltonian of a hydrogen molecule ion, consisting of one electron and two hydrogen nuclei that are separated by a distance $R>0$, is unitarily equivalent to $H(R)+1/R$,
where 
\begin{align*}
H(R):=-\Delta - \dfrac{1}{|x+Re_z/2|}  - \dfrac{1}{|x-Re_z/2|}\,.
\end{align*}
In the following, we are going to work with $H(R)$ instead of $H(R)+1/R$, since the nuclear repulsion term $1/R$ can be added later on without any additional difficulties. First, we note that for the purpose of deriving a lower bound for the ground-state energy of $H(R)$, it suffices to consider $H(R)$ in the subspace of even parity, that is
\begin{align*}
L^2_{\textup{even}}(\R^3):= \left\{\psi \in L^2(\R^3) \; : \; \psi(x)=\psi(-x) \;\textup{a.e.} \right\}\,.   
\end{align*}
Indeed, $H(R)$ commutes with the orthogonal projection $P: L^2(\R^3) \rightarrow L^2_{\textup{even}}(\R^3)$, so the ground state of $H(R)$ has either even or odd parity. However, by the Perron-Frobenius theorem (see e.g. \cite[Theorem XIII.46]{RS4}), the ground state of $H(R)$ is non-degenerate and strictly positive, which, of course, is only possible if it has even parity. Therefore, it suffices to consider $H(R)$ on $D(H(R))=H^2_{\textup{even}}(\R^3):=H^2(\R^3) \cap L^2_{\textup{even}}(\R^3)$ henceforth. Let  $E_2(R)$ denote the second lowest eigenvalue of $H(R)$ (note that $H(R)$ has infinitely many discrete eigenvalues below its essential spectrum $\sigma_{\text{ess}}(H(R))=[0,\infty)$). Then, by Temple's inequality (see e.g. \cite[Theorem XIII.5]{RS4}), we have for all $\psi \in H^2_{\textup{even}}(\R^3)$ satisfying $\|\psi\|=1$ and $\Braket{\psi|H(R)\psi}<E_2(R)$ the lower bound
\begin{align} 
 \inf \sigma(H(R)) \geq \Braket{\psi|H(R)\psi} - \dfrac{\|H(R)\psi \|^2 - \Braket{\psi|H(R)\psi}^2}{E_2(R)-\Braket{\psi|H(R)\psi}}\,.  \label{TempleInequality} 
\end{align}
Our goal is to estimate the right-hand side of \eqref{TempleInequality} for a suitably defined ansatz function $\psi=\psi_R$ (see Eq. \eqref{DefpsiR} below). However, as we do not know the exact value of $E_2(R)$, we first need to derive a lower bound for $E_2(R)$:

\begin{lemma} \label{lmsecondeig}
For all $R>0$,
\begin{align*}
E_2(R) \geq \Ea(1,1)=-\frac{1}{4}\,.
\end{align*}    
\end{lemma}

\begin{proof}
First, we note that by the min-max principle,
\begin{align*}
E_2(R)= \inf_{\stackrel{U \subseteq H^2_{\textup{even}}(\R^3)}{\dim U=2}} \sup_{\stackrel{\psi \in U}{\|\psi\|=1}}  \Braket{\psi|H(R)\psi}\,, 
\end{align*}
where the infimum is taken over all two-dimensional subspaces $U$ of $H^2_{\textup{even}}(\R^3)$. Next, we observe that for all $\psi \in H^2_{\textup{even}}(\R^3)$,
\begin{align*}
\Braket{\psi|H(R)\psi} = \Braket{\psi | \widetilde{H}(R)\psi}\,,
\end{align*}
where the operator
\begin{align*}
\widetilde{H}(R):= -\Delta - \dfrac{2}{\left|x+Re_z/2 \right|}
\end{align*}
has domain $D(\widetilde{H}(R))=H^2(\R^3)$. It is well-known that $\widetilde{H}(R)$ has a non-degenerate ground state (which we denote by $\widetilde{\psi}$ in the following) and the corresponding ground-state energy is given by $\inf \sigma(\widetilde{H}(R))=-1$.
Now, let $U$ denote an arbitrary subspace of $H^2_{\textup{even}}(\R^3)$ with $\dim U=2$. Then there exists some $\psi_{\perp} \in U$ with $\|\psi_{\perp}\|=1$ and $\Braket{\widetilde{\psi}| \psi_{\perp}}=0$. Hence, it follows that
\begin{align*}
\sup_{\stackrel{\psi \in U}{\|\psi\|=1}}  \Braket{\psi|H(R)\psi} \geq 
\Braket{\psi_{\perp} | \widetilde{H}(R)\psi_{\perp}} \geq E_2(\widetilde{H}(R))\,,
\end{align*}
where $E_2(\widetilde{H}(R))$ denotes the second lowest eigenvalue of $\widetilde{H}(R)$. But since $E_2(\widetilde{H}(R))=-\frac{1}{4}=\Ea(1,1)$, the desired estimate follows.
\end{proof}

Our ansatz function for Temple's inequality \eqref{TempleInequality} is
\begin{align}
\psi_R(x)= C_R\Big( \psi_0(x+Re_z/2) + \psi_0(x-Re_z/2)\Big)\,, \label{DefpsiR}
\end{align}
where 
\begin{align*}
\psi_0(x)= \dfrac{\exp(-|x|/2)}{\sqrt{8\pi}}    
\end{align*}
denotes the normalized ground state of the hydrogen atom and the normalization constant $C_R>0$ is chosen so that $\|\psi_R\|=1$. It can be computed explicitly, but its value will cancel out in our calculation. 
We start with the computation of the expectation value $\Braket{\psi_R|H(R)\psi_R}$. First, observe that due to symmetry, 
$$\Braket{\psi_R|H(R)\psi_R}=2C_R\Braket{\psi_R|H(R)\psi_0(x+Re_z/2)}\,.$$
Using now that 
\begin{align*}
H(R)\psi_0(x+Re_z/2)=-\frac{1}{4}\psi_0(x+Re_z/2)-\frac{1}{|x-Re_z/2|}\psi_0(x+Re_z/2)\,,    
\end{align*}
we find
\begin{equation}\label{expcomp}
\Braket{\psi_R|H(R)\psi_R}=-\frac{1}{4}-2(C_R)^2 f(R),
\end{equation}
where 
\begin{equation}\label{fRdef}
f(R):=\Braket{\psi_0(x+Re_z/2) + \psi_0(x-Re_z/2)| \frac{1}{\left|x-Re_z/2\right|} \psi_0(x+Re_z/2) }\,.
\end{equation}
Using spherical coordinates, an elementary but lengthy computation gives
\begin{align*}
\Braket{\psi_0(x+Re_z/2)  | \frac{1}{\left|x-Re_z/2\right|} \psi_0(x+Re_z/2) }= \frac{1}{R}(1-\exp(-R)) - \frac{1}{2} \exp(-R)   
\end{align*}
 and 
 \begin{align*}
 \Braket{ \psi_0(x-Re_z/2) | \frac{1}{\left|x-Re_z/2\right|} \psi_0(x+Re_z/2)}=\left(\frac{1}{2} + \frac{R}{4}\right)   \exp\left(-\frac{R}{2}\right)\,,
 \end{align*}
 so that
\begin{equation}\label{fReqineq}
 f(R)=\frac{1}{R}(1-\exp(-R))-\frac{1}{2} \exp(-R) + \left(\frac{1}{2} + \frac{R}{4}\right) \exp\left(-\frac{R}{2}\right) > 0.
\end{equation}
To estimate the  $L^2$-norm of $H(R) \psi_R$, we first invoke the eigenvalue equation of the ground state $\psi_0$ of the hydrogen atom to obtain 
\begin{align*}
H_R \psi_R=-\frac{1}{4} \psi_R -C_R\left( \frac{ \psi_0(x-Re_z/2)}{|x+Re_z/2|}+\frac{\psi_0(x+Re_z/2)}{|x-Re_z/2|}\right) \,.    
\end{align*}
Consequently, using that $\|\psi_R\|=1$ and recalling Eqs. \eqref{expcomp} and \eqref{fRdef}, we find
\begin{eqnarray}
&\|H(R) \psi_R \|^2= \dfrac{1}{16} + (C_R)^2 (f(R)+g(R))\,,
\label{L2comp}
\end{eqnarray}
where 
\begin{align}
    g(R):=&\left\|\frac{ \psi_0(x-Re_z/2)}{|x+Re_z/2|}+\frac{\psi_0(x+Re_z/2)}{|x-Re_z/2|}\right\|^2 \nonumber \\ = & 2 \left\|\frac{ \psi_0(x-Re_z/2)}{|x+Re_z/2|}\right\|^2 + 2 \Braket{\frac{\psi_0(x+Re_z/2)}{\left|x-Re_z/2\right|}|\frac{\psi_0(x-Re_z/2)}{\left|x+Re_z/2\right|}}.  \label{gRdef}
\end{align}
Using again spherical coordinates, an elementary but lengthy computation gives 
\begin{align}
 2\Braket{\frac{\psi_0(x+Re_z/2)}{\left|x-Re_z/2\right|}|\frac{\psi_0(x-Re_z/2)}{\left|x+Re_z/2\right|}}=\exp\left(-\frac{R}{2}\right) 
 \label{g1comp}
\end{align}
and
\begin{align}
 2 \left\|\frac{\psi_0(x-Re_z/2)}{|x+Re_z/2|}\right\|^2=\frac{R}{2}I(R)\,,   
 \label{g2comp}
\end{align}
 where
\begin{align}\label{IRdef}
 I(R):=\int\limits_0^{\infty} u \ln\left( \dfrac{u+1}{|u-1|} \right)\exp(-Ru) \, du\,.  
\end{align}
In Appendix \ref{Appendix IR} we explain how one can approximate the integral $I(R)$ to arbitrary precision.
Combining Eqs. \eqref{gRdef}, \eqref{g1comp} and \eqref{g2comp}, we arrive at
\begin{equation}\label{gReq}
    g(R)=\exp\left(-\frac{R}{2}\right)+\frac{R}{2}I(R).
\end{equation}
Moreover, it follows from Eqs. \eqref{expcomp}, \eqref{fReqineq} and Lemma \ref{lmsecondeig} that the necessary condition for Temple's inequality \eqref{TempleInequality}, namely  the inequality $\Braket{\psi_R|H(R)\psi_R}<E_2(R)$,  is fulfilled. By Lemma \ref{lmsecondeig}, we can therefore replace  $E_2(R)$ in Temple's inequality \eqref{TempleInequality} with $-\frac{1}{4}$, and after writing everything as a single fraction, we obtain
\begin{equation}
 \inf \sigma(H(R)) \geq   \dfrac{ \frac{1}{4}\Braket{\psi_R|H(R)\psi_R}+\|H(R)\psi_R \|^2}{\frac{1}{4}+\Braket{\psi_R |H(R)\psi_R}}\,.  \label{TempleInequality1} 
\end{equation}
After inserting Eqs. \eqref{expcomp} and \eqref{L2comp} into Eq. \eqref{TempleInequality1}, we arrive at
\begin{equation}
 \inf \sigma(H(R)) \geq   \dfrac{ \frac{f(R)}{2} +g(R)}{-2f(R)}=-\frac{1}{4}-\frac{g(R)}{2f(R)}\,.  \label{TempleInequality2} 
\end{equation}
Next, we insert Eqs. \eqref{fReqineq} and \eqref{gReq} into Eq. \eqref{TempleInequality2}
to obtain  
\begin{align}
\inf \sigma(H(R)) \geq -\dfrac{1}{4} - \dfrac{\dfrac{R}{2} I(R) + \exp\left(-\dfrac{R}{2}\right)}{
\dfrac{2}{R} + \left(1+ \dfrac{R}{2} \right) \exp\left(-\dfrac{R}{2}\right) -\left(1 + \dfrac{2}{R}\right)\exp(-R)}\,.  \label{H2+lower}
\end{align}
To avoid the use of numerical methods in the evaluation of the unknown integral $I(R)$ in Estimate \eqref{H2+lower}, we use the upper bound $I(R) \leq I_K(R)$ for $2K \geq  R$ that is derived in Lemma \ref{lmI(R)Upper} in the appendix, where $I_K(R)$ is defined in Eq. \eqref{DefIKR} below. Since the denominator in Estimate \eqref{H2+lower} is positive for all $R>0$, we arrive at
\begin{align}
\inf \sigma(H(R)) \geq -\dfrac{1}{4} - \dfrac{\dfrac{R}{2} I_K(R) + \exp\left(-\dfrac{R}{2}\right)}{
\dfrac{2}{R} + \left(1+ \dfrac{R}{2} \right) \exp\left(-\dfrac{R}{2}\right) -\left(1+\dfrac{2}{R} \right)\exp(-R)} \qquad \quad (2K \geq R)\,.  \label{H2+lowerK}
\end{align}

%\paragraph{Proof of Theorem \ref{Thm:EnergyCondition}}

In view of Eqs. \eqref{DefEinfinity} and \eqref{EqEinfinity}, we have the rough upper bound
\begin{align*}
 E(2,(1,1)) \leq E(2,(1,1),\infty) \leq 2 \Ea(1,1)\,,
\end{align*}
where $\Ea(1,1)=-1/4$. Hence, to prove Theorem \ref{Thm:EnergyCondition}, it suffices to show that
\begin{align}
E(1,(1,1)) > -\dfrac{31}{64}\,.
\end{align}
For this purpose, we first note that the bound \eqref{Rbestimatediatomicexplicit} implies that the binding length of $\ce{H2+}$ has the lower bound $R_1 \geq 2$, so we actually have to show that
\begin{align}
E(1,(1,1))=\inf_{R\geq 2}\left( \inf\sigma(H(R)) + \dfrac{1}{R}\right) > -\dfrac{31}{64}\,. \label{Eesttoproveweak}
\end{align}
In fact, we are going to establish a slightly stronger estimate, namely that for all $R\geq 2$,
\begin{align}
\inf\sigma(H(R)) + \dfrac{1}{R} >  -0.4797\,.  \label{Eesttoprove}
\end{align}
To this end, we use the well-known fact that $\inf\sigma(H(R))$ is monotonically increasing as a function of $R>0$ (see e.g. \cite[Theorem 4.6.20 (i)]{Thirring}). Therefore, Estimate \eqref{Eesttoprove} on a given interval $[a,b)$ with $a<b$ would follow from
\begin{align*}
\inf\sigma(H(a)) + \dfrac{1}{b} > -0.4797\,.    
\end{align*}
This observation allows us to reduce the proof of Estimate \eqref{Eesttoprove} to finitely many subintervals $I_j$, where on each subinterval $I_j=[a_j,b_j)$ we only have to evaluate Estimate \eqref{H2+lowerK}  in the left endpoint $R=a_j$. We choose the partition $[2,\infty)=I_1 \cup I_2 \cup I_3$, where $I_1=[2,3.3)$, $I_2=[3.3,6)$ and 
$I_3=[6,\infty)$. Then, using Estimate \eqref{H2+lowerK} with $K=10$ together with \eqref{DefIKR} and \eqref{gammaineq} from Appendix \ref{Appendix IR}, we find that\footnote{If one only considers four decimal places, then Estimate \eqref{intervalproof} is also  the sharpest estimate that one can obtain from precise numerical calculations of $I(2), I(3.3)$ and $I(6)$.} 
\begin{align}
\min\left(\inf\sigma(H(2)) + \dfrac{1}{3.3},\; \inf\sigma(H(3.3)) + \dfrac{1}{6},\; \inf\sigma(H(6))\right)>-0.4797\,. \label{intervalproof}
\end{align}
 This establishes Estimate \eqref{Eesttoprove} on each of the subintervals $I_1, I_2, I_3$, and hence Estimate \eqref{Eesttoprove} is valid on the whole interval $[2,\infty)$.\hfill$\Box$
\begin{remark}
    By dividing the interval $[2,\infty)$ into more subintervals, it is possible to improve the gap in Estimate \eqref{Eesttoprove} and, consequently, to improve Theorem \ref{Thm:EnergyCondition}. 
\end{remark}

\section*{Acknowledgements}
The research for this article was 
funded by the Deutsche Forschungsgemeinschaft (DFG, German Research Foundation) – Project-ID 258734477 – SFB 1173. 

\appendix
\section*{Appendix}
\setcounter{section}{0}\setcounter{theorem}{0}
\numberwithin{theorem}{section}
\section{Proof of Estimate \texorpdfstring{\eqref{EUpper}}{(27)}}\label{Appendix Upper}
\begin{lemma} \label{lmUpperproof}
For all $Z \geq \frac{5}{16}$, the ground-state energy $\Ea(2,Z)$ defined by Eqs. \eqref{DefEat} and \eqref{Eateta=1} satisfies the upper bound
\begin{align}
\Ea(2,Z) \leq -\frac{1}{2}\left(Z-\frac{5}{16} \right)^2. \label{EUpperNeu}
\end{align}
\end{lemma}

\begin{proof}
For $Z=\frac{5}{16}$, Estimate \eqref{EUpperNeu} follows from the fact that $\Ea(2,Z)<0$ for all $Z>0$, so it suffices to consider the case $Z>\frac{5}{16}$ only. To this end, we first note that the Hamiltonian $\Ha(2,Z)$ defined by Eq. \eqref{DefHat}
with $\eta=1$ acts trivially in the spin variables. Hence, by the variational principle, we have for any normalized $\psi \in H^2(\R^3)$ the upper bound
\begin{align*}
\Ea(2,Z) \leq \Braket{\psi \otimes \psi | \Ha(2,Z) \psi \otimes \psi} = 
2\| \nabla \psi \|^2 - 2Z \Braket{\psi | \dfrac{1}{\left|x\right|} \psi} + \Braket{\psi \otimes \psi | \dfrac{1}{\left|x_1-x_2\right|} \psi \otimes \psi}\,.
\end{align*}
We make the ansatz $\psi=\psi_{\lambda}$, where
\begin{align*}
\psi_{\lambda}(x)= \sqrt{\frac{\lambda^3}{8\pi}} \exp(-\lambda \left| x \right|/2)  
\qquad \qquad (\lambda>0) \end{align*}
denotes the normalized ground state of the one-particle operator $-\Delta- \frac{\lambda}{\left| x \right|}$. Then, a straightforward computation yields that
$\| \nabla \psi \|^2 = \lambda^2/4$, $\Braket{\psi | \dfrac{1}{\left|x\right|} \psi} = \lambda/2$, and, by Newton's theorem,
\begin{align*}
\Braket{\psi \otimes \psi | \dfrac{1}{\left|x_1-x_2\right|} \psi \otimes \psi}
=\frac{\lambda^6}{4} \int_{0}^{\infty}\int_{0}^{\infty} \dfrac{\exp(-\lambda r_1)\exp(-\lambda r_2)}{\max(r_1,r_2)} (r_1)^2 (r_2)^2  \, dr_2\, dr_1 = \frac{5}{16} \lambda.
\end{align*}
We conclude that for all $\lambda>0$,
\begin{align*}
\Ea(2,Z) \leq \frac{\lambda^2}{2} - Z\lambda + \frac{5}{16} \lambda\,.
\end{align*}
Estimate \eqref{EUpperNeu} now follows by optimizing in $\lambda>0$ (the optimal choice is $\lambda=Z-\frac{5}{16}$).  
\end{proof}

\section{Analysis of the integral \texorpdfstring{$I(R)$}{I(R)} from Eq. \texorpdfstring{\eqref{IRdef}}{(53)}} \label{Appendix IR}
Recall that the exponential integral function $\textup{Ei}(\cdot)$ is defined for $x \neq 0$ as 
\begin{equation}\label{Eidef}
\textup{Ei}(x)=-\int_{-x}^\infty \frac{e^{-v}}{v}\,dv=\int_{-\infty}^x \frac{e^{v}}{v}\,dv,
\end{equation}
where for $x>0$ the integrals are understood in the sense of Cauchy principal values. For convenience of the reader, we outline the proof of the following well-known lemma:
\begin{lemma}
\begin{align}
\textup{Ei}(x)= \gamma + \ln(|x|) + \sum\limits_{n=1}^{\infty} \dfrac{x^n}{n \cdot n!}   \qquad \qquad (x \neq 0)\,, \label{Eiseries}
\end{align}
where $\gamma=-\int_{-\infty}^0 \ln(|v|) e^{v}\;dv$ denotes the Euler-Mascheroni constant, see e.g. \cite[Eq. (2.2.8)]{Lagarias}.
\end{lemma}
\begin{proof}
In the sense of Cauchy principal values we  have
\begin{equation*}
\text{Ei}(x)= \int_{-\infty}^1 \frac{e^{v}}{v}\;dv + \int_1^x \frac{1}{v}\;dv + \int_0^x \frac{e^v-1}{v}\;dv-\int_0^1 \frac{e^v-1}{v}\;dv
\end{equation*}
or
\begin{equation}\label{Ei1}
\text{Ei}(x)=\int_{-\infty}^1 \frac{e^{v}}{v}\;dv -\int_0^1 \frac{e^v-1}{v}\;dv + \ln(|x|) + \int_0^x \frac{e^v-1}{v}\;dv.
\end{equation}
Using the Taylor series of $e^v$ and the dominated convergence theorem, we find that 
\begin{equation}\label{Ei2}
\int_0^x \frac{e^v-1}{v}\;dv= \sum\limits_{n=1}^{\infty} \dfrac{x^n}{n \cdot n!}\,.
\end{equation}
By the definition of the Cauchy principal value and integration by parts, we obtain that
\begin{equation}\label{Ei3}
 \int_{-\infty}^1 \frac{e^{v}}{v}\;dv = -\int_{-\infty}^0 \ln(|v|) e^v\;dv - \int_0^1  \ln(v) e^v\;dv.
\end{equation}
We also observe that
\begin{align*}
\ln(v) e^v + \frac{e^v-1}{v}= [\ln(v) (e^v-1)]'\,,
\end{align*}
which implies that
\begin{equation}\label{Ei4}
\int_0^1  \ln(v) e^v\;dv + \int_0^1 \frac{e^v-1}{v}\;dv = 0\,.
\end{equation}
Combining Eqs. \eqref{Ei1}, \eqref{Ei2}, \eqref{Ei3}, \eqref{Ei4}, we arrive at Eq. \eqref{Eiseries}.
\end{proof}
We have the following representation of the integral $I(R)$:
\begin{lemma}\label{lmI(R)}
    Let $I(R)$ be defined by Eq. \eqref{IRdef}. Then for all $R>0$ we have
    \begin{equation}\label{IRvalue}
I(R)= \left(\frac{1}{R}-\frac{1}{R^2}\right) e^R \textup{Ei}(-R) + \left(\frac{1}{R^2}+\frac{1}{R}\right) e^{-R} \textup{Ei}(R),
    \end{equation}
    where the last integral is understood in the sense of the Cauchy principal value. 
\end{lemma}
\begin{proof}
We use the decomposition
\begin{equation}\label{IRdec}
I(R)=I_1-(I_2+I_3),
\end{equation}
where
\begin{align*}
I_1=\int_0^\infty \ln(t+1)t\exp(-Rt)\;dt\,, \\ 
I_2=\int_0^1 \ln(1-t)t\exp(-Rt)\;dt\,, \\
I_3=\int_1^\infty \ln(t-1)t\exp(-Rt)\;dt\,.
\end{align*}
Observing that $\big[\left(\frac{-t}{R}-\frac{1}{R^2}\right)\exp(-Rt)\big]'=t\exp(-Rt)$, we can apply integration by parts to obtain that
\begin{equation*}
I_1=\frac{1}{R^2} +\left(\frac{1}{R^2}-\frac{1}{R}\right) \int_0^\infty \frac{\exp(-Rt)}{t+1}\;dt\,,
\end{equation*}
which after the substitution $v=R(t+1)$ gives
\begin{equation}\label{I1value}
I_1=\frac{1}{R^2} +\left(\frac{1}{R^2}-\frac{1}{R}\right) e^R \int_R^\infty \frac{e^{-v}}{v}\;dv.
\end{equation}
Writing $I_2=\lim_{\varepsilon \to 0+}\int_0^{1-\varepsilon} \ln(1-t)t\exp(-Rt) \,dt$, an integration by parts similar to the above one yields
\begin{equation*}
I_2=\int_0^1\frac{\exp(-Rt)}{R}\;dt + \left(\frac{1}{R^2}+\frac{1}{R}\right)\lim_{\varepsilon \to 0+}\left( \int_0^{1-\varepsilon} \frac{\exp(-Rt)}{t-1}\;dt-\ln(\varepsilon)e^{-R}\right)\,,
\end{equation*}
where we have also used that $\lim_{\varepsilon \to 0+} \varepsilon \ln(\varepsilon)=0$. With the help of the substitution $v=R(t-1)$, we arrive at
\begin{equation}\label{I2value}
I_2=\int_0^1\frac{\exp(-Rt)}{R}\;dt + \left(\frac{1}{R^2}+\frac{1}{R}\right) e^{-R}\lim_{\varepsilon \to 0+}\left( \int_{-R}^{-\varepsilon R} \frac{e^{-v}}{v}\;dv-\ln(\varepsilon)\right)\,.
\end{equation}
Completely analogously we find
\begin{equation}\label{I3value}
I_3=\int_1^{\infty}\frac{\exp(-Rt)}{R}\;dt + \left(\frac{1}{R^2}+\frac{1}{R}\right) e^{-R}\lim_{\varepsilon \to 0+}\left( \int_{\varepsilon R}^\infty \frac{e^{-v}}{v}\;dv+\ln(\varepsilon)\right)\,.
\end{equation}
Adding Eqs. \eqref{I2value} and \eqref{I3value}, we obtain that
\begin{equation}\label{I2plusI3}
I_2+I_3=\frac{1}{R^2}+ \left(\frac{1}{R^2}+\frac{1}{R}\right) e^{-R} \int_{-R}^\infty \frac{e^{-v}}{v}\,dv\,,
\end{equation}
where the integral is understood in the sense of the Cauchy principal value. Using Eq. \eqref{IRdec} together with Eqs. \eqref{Eidef}, \eqref{I1value} and \eqref{I2plusI3}, we arrive at Eq. \eqref{IRvalue}.
\end{proof}

\begin{lemma}\label{lmI(R)Upper}
 Let $R \geq 1$ be given and let $I(R)$ be defined by Eq. \eqref{IRdef}. For $K \in \N$ with $2K \geq R$ we define
 \begin{align}
 I_K(R)&:= \left(\frac{1}{R}-\frac{1}{R^2}\right) e^R
 \left(\gamma + \ln(R) +  \sum_{k=1}^{ 2K } \dfrac{(-R)^k}{k\cdot k!} \right) \nonumber \\
 &\quad\,+  \left(\frac{1}{R^2}+\frac{1}{R}\right) e^{-R} \left(\gamma + \ln(R) + \sum_{k=1}^{2K} \dfrac{R^k}{k\cdot k!}  + \dfrac{R^{2K+1}e^R}{(2K+1)(2K+1)!} \right)\,. \label{DefIKR}
 \end{align}
 Then 
 \begin{align}
 0 \leq  I_K(R) - I(R) \leq  \dfrac{(R^{2K}-R^{2K-1}) e^R}{(2K+1)(2K+1)!}   +  \dfrac{R^{2K} + R^{2K-1}}{(2K+1)(2K+1)!}\,.
 \label{Iremainder2}
 \end{align}
 In particular, $\lim_{K \to \infty} I_K(R)=I(R)$.
\end{lemma}
\begin{remark}
The Euler-Mascheroni constant $\gamma$ appearing in Eq. \eqref{DefIKR} satisfies the inequality
\begin{equation}\label{gammaineq}
\gamma \leq 0.57722.
\end{equation}
 We refer e.g. to \cite{Lagarias}, where the first $50$ digits of $\gamma$ are computed. 
\end{remark}
\begin{proof}
We use Lemma \ref{lmI(R)} together with the series expansion from Eq. \eqref{Eiseries} for the exponential integral function. Note that
\begin{align*}
 a_k:= \dfrac{R^k}{k \cdot k!}   
\end{align*}
is monotonically decreasing for $k \geq R$ with $\lim_{k \to \infty} a_k=0$. Since  $2K \geq R$ by assumption, it follows from Eq. \eqref{Eiseries} and the Leibniz remainder estimate that
 \begin{align}
 0 \leq   \left(\gamma + \ln(R) + \sum\limits_{k=1}^{2K} \dfrac{(-R)^k}{k \cdot k!} \right)-    \textup{Ei}(-R) \leq \dfrac{R^{2K+1}}{(2K+1) \cdot (2K+1)!} \,. \label{FirstEiEst2} 
 \end{align}
Similarly, for $\textup{Ei}(R)$ with $R \geq 1$, we find that 
\begin{align}
\gamma + \ln(R) + \sum\limits_{k=1}^{2K} \dfrac{R^k}{k \cdot k!} \leq \textup{Ei}(R)&= \gamma + \ln(R) + \sum\limits_{k=1}^{\infty} \dfrac{R^k}{k \cdot k!} \nonumber \\
&\leq  \gamma + \ln(R) + \sum\limits_{k=1}^{2K} \dfrac{R^k}{k\cdot k!} + \dfrac{R^{2K+1}}{(2K+1)(2K+1)!} \sum\limits_{k=2K+1}^{\infty} \dfrac{R^{k-2K-1}}{(k-2K-1)!} \nonumber \\
& = \gamma + \ln(R) + \sum\limits_{k=1}^{2K} \dfrac{R^k}{k \cdot k!} + \dfrac{R^{2K+1}e^{R}}{(2K+1)(2K+1)!}\,. \label{SecondEi2}
\end{align}
Estimate \eqref{Iremainder2} follows from Eq. \eqref{IRvalue} by applying Estimates \eqref{FirstEiEst2} and \eqref{SecondEi2}.
\end{proof}

\bibliographystyle{plain}
\bibliography{LowerBoundsBib}

@String {cmp="Comm. Math. Phys."}

@String {cpam="Commun. pure appl. math."}

@String {ijqc="Int. J. Quantum Chem."}

@String {jmp="J. Math. Phys."}

@String {jpa="J. Phys. A"}

@String {jpca="J. Phys. Chem. A"}

@String {lmp="Lett. Math. Phys."}

@String {pra="Phys. Rev. A"}

@String {prl="Phys. Rev. Lett."}

@article{Cornu_2009,
doi = {10.1088/1751-8113/42/49/495001},
url = {https://doi.org/10.1088/1751-8113/42/49/495001},
year = {2009},
publisher = {},
volume = {42},
number = {49},
pages = {495001},
author = {F. Cornu and P.~A. Martin},
title = {Atom–wall dispersive forces: a microscopic approach},
journal = jpa
}

@article {ABH,
    AUTHOR = {Anapolitanos, I. and Badalyan, M. and Hundertmark,
              D.},
     TITLE = {On the van der {W}aals interaction between a molecule and a
              half-infinite plate},
   JOURNAL = jmp,
  FJOURNAL = {Journal of Mathematical Physics},
    VOLUME = {64},
      YEAR = {2023},
    NUMBER = {8},
     PAGES = {Paper No. 082103, 28},
      ISSN = {0022-2488,1089-7658},
   MRCLASS = {81V55 (81Q10)},
  MRNUMBER = {4627083},
MRREVIEWER = {Francisco\ Gonzalez Montoya},
       DOI = {10.1063/5.0117054},
       URL = {https://doi.org/10.1063/5.0117054},
}

@article{LewinConvexityN,
  title = {Ground {S}tate {E}nergy {I}s {N}ot {A}lways {C}onvex in the {N}umber of {E}lectrons},
  author = {Di Marino, S. and Lewin, M. and Nenna, L.},
  journal = jpca,
  volume = {128},
  issue = {49},
  pages = {10697--10706},
  year = {2024}
  }

@misc{LowerBoundsAtoms,
  author        = {M. Hofacker},
  title         = {{R}igorous {L}ower {B}ounds for {A}tomic and {I}onic {G}round-{S}tate {E}nergies},
  year          = {2026},
  eprint        = {2610.34480},
  archivePrefix = {arXiv},
  primaryClass  = {math-ph},
  howpublished = {\url{https://arxiv.org/abs/2610.34480}}
}

@article{LiebThirringvdW,
  title = {Universal nature of van der {W}aals forces for {C}oulomb systems},
  author = {Lieb, E.~H. and Thirring, W.~E.},
  journal = prl,
  volume = {34},
  number = {1},
  pages = {40--46},
  year = {1986},
  }

@article {MorganSimon,
    AUTHOR = {Morgan, J.~D. and Simon, B.},
     TITLE = {{B}ehavior of {M}olecular {P}otential {E}nergy {C}urves for {L}arge {N}uclear {S}eparations},
   JOURNAL = ijqc,
  FJOURNAL = {},
    VOLUME = {17},
      YEAR = {1980},
    NUMBER = {6},
     PAGES = {1143-1166},
      ISSN = {},
   MRCLASS = {},
  MRNUMBER = {},
       DOI = {},
       URL = {},
}

@article {Briet,
    AUTHOR = {Briet, P. and Duclos, P. and Hogreve, H.},
     TITLE = {{M}onotonicity of {I}onization {E}nergies for {T}hree {E}lectron {L}ithium-{T}ype {S}ystems},
   JOURNAL = lmp,
  FJOURNAL = {Letters in Mathematical Physics. A Journal for the Rapid Dissemination of Short Contributions in the Field of
              Mathematical Physics},
    VOLUME = {13},
      YEAR = {1987},
    NUMBER = {2},
     PAGES = {137--140},
      ISSN = {0377-9017},
   MRCLASS = {81G45},
  MRNUMBER = {886149},
       DOI = {10.1007/BF00955202},
       URL = {https://doi.org/10.1007/BF00955202},
}

@book{Thirring,
  title = {Quantum Mathematical Physics:  Atoms, Molecules and  Large  Systems},
  edition ={2nd},
  author = {W. Thirring},
  year= {2002}, 
  publisher = {Springer (Berlin, Heidelberg)},
}

@article {SolovejMolecules,
    AUTHOR = {Solovej, J. P.},
     TITLE = {Asymptotic {N}eutrality of {D}iatomic 
     {M}olecules},
   JOURNAL = cmp,
  FJOURNAL = {Communications in Mathematical Physics},
    VOLUME = {130},
      YEAR = {1990},
    NUMBER = {1},
     PAGES = {185--204},
      ISSN = {0010-3616,1432-0916},
   MRCLASS = {81V55},
  MRNUMBER = {1055692},
MRREVIEWER = {C.\ R.\ Garibotti},
       URL = {http://projecteuclid.org/euclid.cmp/1104187935},
}

@article {MoleculesBenguria,
    AUTHOR = {Alarc\'{o}n, H.~R. and Benguria, R.~D.},
     TITLE = {A {L}ower {B}ound on the {S}ize of {M}olecules},
   JOURNAL = lmp,
  FJOURNAL = {Letters in Mathematical Physics},
    VOLUME = {35},
      YEAR = {1995},
    NUMBER = {3},
     PAGES = {281--289},
      ISSN = {0377-9017,1573-0530},
   MRCLASS = {81V55 (81Q05)},
  MRNUMBER = {1352397},
MRREVIEWER = {Mary\ Beth\ Ruskai},
       DOI = {10.1007/BF00761300},
       URL = {https://doi.org/10.1007/BF00761300},
}

@article{AnapolitanosSigal2013,
      title={Long {R}ange {B}ehaviour of van der {W}aals Force}, 
      author={Anapolitanos, I. and Sigal, I.~M.},
   JOURNAL = cpam,
    VOLUME = {70},
      YEAR = {2017},
    NUMBER = {9},
     PAGES = {1633--1671},
       DOI = {10.1002/cpa.21695},
       URL = {https://doi.org/10.1002/cpa.21695},
}

@article{Lagarias,
  title = {Euler’s constant: Euler’s work and modern developments
},
  author = {Lagarias, J.~C.},
  journal = {Bull. Amer. Math. Soc.},
 volume = {50},
  pages = {527--628},
  numpages = {0},
  year = {2013},
  month = { },
  publisher = {American Mathematical Society},
  doi = {10.1090/S0273-0979-2013-01423-X},
  url = {https://doi.org/10.1090/S0273-0979-2013-01423-X}
}

@article{Anap16,
  title = {Remainder {E}stimates for the {L}ong {R}ange {B}ehavior of the van der {W}aals {I}nteraction {E}nergy},
  author = {Anapolitanos, I.},
  journal = {Ann. Henri Poincaré},
 volume = {17},
  pages = {1209-1261},
  year = {2016},
  doi = {10.1007/s00023-015-0437-6},
  url = {https://doi.org/10.1007/s00023-015-0437-6}
}

@article{AnOlZa,
  title = {On boundedness of isomerization paths for non- and semirelativistic molecules},
  author = {Anapolitanos, I. and Olivieri, M. and Zalczer, S.},
  journal = {J. Funct. Anal.},
 volume = {288},
   issue = {3},
  number = {110713},
  year = {2025},
  doi = {10.1007/s00205-019-01475-5},
  url = {https://doi.org/10.1007/s00205-019-01475-5}
}

@article{AnapLew,
  title = {Compactness of {M}olecular {R}eaction {P}aths in {Q}uantum {M}echanics},
  author = {Anapolitanos, I. and Lewin, M.},
  journal = {Arch. Ration. Mech. Anal.},
 volume = {236},
   issue = {2},
  pages = {505-576},
  year = {2020},
  doi = {10.1007/s00205-019-01475-5},
  url = {https://doi.org/10.1007/s00205-019-01475-5}
}

@article{BaHaHuVu,
  title = {Van der {W}aals–{L}ondon interaction of atoms with pseudorelativistic kinetic energy},
  author = {Barbaroux, J.~M. and Hartig, M. and Hundertmark, D. and Vugalter, S.},
  journal = {Analysis and PDE},
 volume = {15},
   issue = {5},
  pages = {1375-1428},
  year = {2022},
  doi = {10.2140/apde.2022.15.1375},
  url = {https://doi.org/10.2140/apde.2022.15.1375}
}

@article{CaSc,
  title = {Van der {W}aals {I}nteractions {B}etween {T}wo {H}ydrogen {A}toms: {T}he {S}later--{K}irkwood {M}ethod Revisited},
  author = {Cances, E. and Scott, R.},
  journal = {SIAM J. Math. Anal.},
 volume = {50},
   issue = {1},
  pages = {381--410},
  year = {2018},
  doi = {10.1137/15M1021878},
  url = {https://doi.org/10.1137/15M1021878}
}

@article{CaCoSc,
  title = {{V}an der {W}aals {I}nteractions {B}etween {T}wo {H}ydrogen {A}toms: {T}he {S}later--{K}irkwood {M}ethod {R}evisited},
  author = {Cances, E. and  Coyaud, R. and Scott, R.},
  journal = {Commun. Math. Sci.},
 volume = {21},
   issue = {4},
  pages = {915--948},
  year = {2023},
  doi = {10.4310/CMS.2023.v21.n4.a2},
  url = {https://doi.org/10.4310/CMS.2023.v21.n4.a2}
}

@article{LiebIonization,
  title = {Bound on the maximum negative ionization of atoms and molecules},
  author = {Lieb, E.~H.},
  journal = pra,
  volume = {29},
  issue = {6},
  pages = {3018--3028},
  numpages = {0},
  year = {1984},
  month = { },
  publisher = {American Physical Society},
  doi = {10.1103/PhysRevA.29.3018},
  url = {https://link.aps.org/doi/10.1103/PhysRevA.29.3018}
}

@book {Teschl,
    AUTHOR = {Teschl, G.},
     TITLE = {Mathematical Methods in Quantum Mechanics},
    SERIES = {Graduate Studies in Mathematics},
    VOLUME = {99},
      NOTE = {With Applications to Schr\"{o}dinger Operators},
 PUBLISHER = {American Mathematical Society, Providence, RI},
      YEAR = {2009},
     PAGES = {xiv+305},
      ISBN = {978-0-8218-4660-5},
   MRCLASS = {81-01 (47N50 81Qxx)},
  MRNUMBER = {2499016},
MRREVIEWER = {Rupert L. Frank},
       DOI = {10.1090/gsm/099},
       URL = {https://doi.org/10.1090/gsm/099},
}

@book {RS4,
    AUTHOR = {Reed, M. and Simon, B.},
     TITLE = {Methods of modern mathematical physics. {IV}. {A}nalysis of
              operators},
 PUBLISHER = {Academic Press [Harcourt Brace Jovanovich, Publishers], New
              York-London},
      YEAR = {1978},
     PAGES = {xv+396},
      ISBN = {0-12-585004-2}
}

\end{document}